\documentclass[12pt,fleqn]{article}%

\usepackage{amsmath,amssymb,amsthm,enumerate,color,url,adjustbox}

\usepackage[pdftex,
colorlinks,%
linkcolor=blue,citecolor=red,urlcolor=blue,
hyperindex,%
plainpages=false,%
bookmarksopen,%
bookmarksnumbered,%
unicode]{hyperref}

\usepackage{cmap}
\usepackage[T1]{fontenc}
\usepackage{lmodern}
\usepackage[utf8]{inputenc}
\usepackage[english]{babel}
\usepackage{array,longtable}

\newcommand{\Int}{\mathop{\rm Int}\nolimits}

\theoremstyle{definition}

\newtheorem{Example}{Example}

\begin{document}

\begin{center}
\Large\bfseries Invariants of seven-dimensional nilpotent Lie algebras
\end{center}

\par\noindent{\large
Vyacheslav M.\ Boyko$^{\dag\ddag}$ and Kristina Yu.\ Leleka$^{\S}$}

\vspace{4mm}

\noindent{\small\itshape
$^\dag$~Institute of Mathematics of NAS of Ukraine, Kyiv, Ukraine}

\vspace{2mm}

\noindent{\small\itshape
$^\ddag$~Department of Mathematics, Kyiv Academic University, Kyiv, Ukraine}

\vspace{2mm}

\noindent{\small\itshape
$^\S$~Taras Shevchenko National University of Kyiv, Ukraine}

\vspace{5mm}

\noindent{\small email: \href{mailto:boyko@imath.kyiv.ua}{boyko@imath.kyiv.ua}, \href{mailto:kristinaleleka18@gmail.com}{kristinaleleka18@gmail.com}}

\begin{abstract}
Using the method of moving frames, we construct polynomial functional bases of invariants for all seven-dimensional complex nilpotent Lie algebras in Seeley's classification as corrected by Gong. The use of~$K$-canonical bases simplifies the successive normalization of the components of the fundamental lifted invariant. The procedure is illustrated by examples, and the resulting bases of Casimir operators are listed in a single table. The infinitesimal invariance equations are used for independent verification.
\end{abstract}


\medskip


\section{Introduction}
Invariants of Lie algebras, also known as generalized Casimir operators, are among the principal characteristics of Lie algebras and provide insight into their structure and properties. They play an important role in representation theory, the study of integrability of Hamiltonian systems, quantum mechanics and symmetry analysis.
Of particular interest are polynomial functional bases of invariants of Lie algebras, whose symmetrizations are Casimir operators (see~\cite{Dixmier} for symmetrization and~\cite{Casimir} for the classical construction). The existence of polynomial functional bases for Lie algebras with nilpotent radicals, including nilpotent and perfect Lie algebras, is discussed in~\cite{Abellanas-Martinez,Ndogmo}. This should be distinguished from the stronger property that the algebra of polynomial invariants itself is a polynomial algebra; see~\cite{Ooms2009,Ooms2012}.

The classical approach to the construction of generalized Casimir operators is based on the integration of overdetermined systems of first-order linear partial differential equations~\cite{Abellanas-Martinez,Abellanas-Martinez2,patera-sharp,Pecina-Cruz}. This approach is, however, difficult to implement and rapidly becomes computationally cumbersome even for low-dimensional Lie algebras.

An alternative is the algebraic approach proposed by Boyko, Patera and Popovych~\cite{Boyko2006}, based on Cartan's method of moving frames in the formulation of Fels and Olver~\cite{Fels-Olver,Fels-Olver2}. The extended version of~\cite{Boyko2006} treats real solvable Lie algebras of dimension at most five, real six-dimensional nilpotent Lie algebras and six-dimensional solvable Lie algebras with four-dimensional nilradicals; see also the earlier computations in~\cite{patera-sharp}. The approach was subsequently applied to solvable Lie algebras of arbitrary finite dimension with prescribed structures of their nilradicals~\cite{Boyko2007a} and to triangular Lie algebras and their solvable extensions~\cite{Boyko2007b,Boyko2007c,Boyko2008}. A survey of the method and its applications is given in~\cite{Boyko&Patera&Popovych2009}. Within this approach, invariants are obtained by eliminating group parameters from lifted invariants, avoiding the integration of systems of differential equations.

The structure of the invariants depends directly on the structure of the Lie algebra itself. The general classification of finite-dimensional Lie algebras is an extremely difficult problem and has not been completed in arbitrary dimension. Therefore, studies are usually restricted to particular classes of Lie algebras or to Lie algebras of fixed dimension (see the review in~\cite{popovych2003}).

Nilpotent Lie algebras occupy an important place among various classes of Lie algebras. Their central series make them amenable to constructive methods of investigation.

The invariants in dimension seven have also been studied from the viewpoint of enveloping and symmetric algebras. Ooms~\cite{Ooms2009,Ooms2012} described the centers of the enveloping algebras and the Poisson centers of the symmetric algebras for indecomposable complex nilpotent Lie algebras of dimension at most seven. In particular, these results go beyond the construction of polynomial functional bases, since they describe the algebra of polynomial invariants. Ooms credits Romdhani~\cite{Romdhani} with earlier generators of the invariant fields in dimension seven and discusses omissions and corrections in the introduction to~\cite{Ooms2009} and in Section~3 of~\cite{Ooms2012}. Pecina-Cruz~\cite{Pecina-Cruz} also applied the infinitesimal approach to seven-dimensional nilpotent Lie algebras. For the algebras in Gong's list, Vorushilov~\cite[Section~2]{Vorushilov} tabulates complete sets of polynomials in bi-involution and the sizes of the Jordan and Kronecker blocks.

The aim of this paper is to present polynomial functional bases of invariants for all seven-dimensional complex nilpotent Lie algebras in Seeley's classification as corrected by Gong, using the moving-frame method in~$K$-canonical bases.

Let~$\mathfrak g$ be a finite-dimensional Lie algebra over~$\mathbb R$ or~$\mathbb C$, write~$n=\dim\mathfrak g$, and let~$G$ be a connected Lie group with Lie algebra~$\mathfrak g$. Denote by~$\mathfrak g^*$ the dual space of the vector space~$\mathfrak g$.
The commutation relations between the basis elements~$e_1,\ldots,e_n$ of~$\mathfrak g$ are given by
$[e_i,e_j]=c_{ij}^k e_k$,
where~$c_{ij}^k$ are the structure constants of~$\mathfrak g$. Here and in what follows, summation over repeated indices is understood.
The \emph{coadjoint representation} of~$G$ is the mapping~${\rm Ad}^*\colon G\to {\rm GL}(\mathfrak g^*)$ defined for every~$g\in G$, $x\in\mathfrak g^*$ and~$u\in\mathfrak g$ by
\[
\langle{\rm Ad}^*_g x,u\rangle=\langle x,{\rm Ad}_{g^{-1}}u\rangle.
\]
Here~${\rm Ad}\colon G\to {\rm GL}(\mathfrak g)$ is the adjoint representation of~$G$ on~$\mathfrak g$, and the image~${\rm Ad}_G$ of~$G$ under~${\rm Ad}$ is the group~$\Int(\mathfrak g)$ of inner automorphisms of~$\mathfrak g$.
The \emph{rank} $\operatorname{rank}{\rm Ad}^*_G$ of the coadjoint representation of~$G$ is the maximal dimension of a coadjoint orbit, computed over the ground field.

A smooth function~$F$ in the real case, or a holomorphic function~$F$ in the complex case, is called an \emph{invariant} of~${\rm Ad}^*_G$ if
$F({\rm Ad}_g^* x)=F(x)$
for all~$g\in G$ and~$x\in\Omega$ such that~${\rm Ad}_g^*x\in\Omega$, where~$\Omega$ is a domain in the dual space~$\mathfrak g^*$.
Denote by~${\mathop{\rm Inv}\nolimits}({\rm Ad}^*_G)$ the set of invariants of the group~${\rm Ad}^*_G$ on~$\Omega$.
Near a regular point of~$\mathfrak g^*$, the maximal number of functionally independent local first integrals of the coadjoint orbit foliation is $N_\mathfrak g=\dim\mathfrak g-\operatorname{rank}{\rm Ad}^*_G$; see~\cite{beltrametti,Boyko2006,Salgado-Gonzalez}. Equivalently, $N_\mathfrak g$ is the index of~$\mathfrak g$, given by $\dim\mathfrak g-\operatorname{rank}C(x)$, where~$C(x)=(c_{ij}^kx_k)_{i,j=1}^n$ and the rank is generic~\cite{Ooms2009,Ooms2012}.

Fixing a basis~$\mathcal E=(e_1,\ldots,e_n)$ of the Lie algebra~$\mathfrak g$ also fixes the dual basis~$\mathcal E^*=(e_1^*,\ldots,e_n^*)$ of~$\mathfrak g^*$ and hence gives rise to matrix representations of~${\rm Int}(\mathfrak g)$ and~${\rm Ad}^*_G$.
Write~$x=\sum_{i=1}^n x_i e_i^*$ and let~$\check x=(x_1,\ldots,x_n)$ be its coordinate row vector. In these coordinates the coadjoint action of~$g$ is right multiplication by the matrix of~$\operatorname{Ad}_{g^{-1}}$ in~$\mathcal E$.

It is well known that there is a one-to-one correspondence between the elements of the center of the universal enveloping algebra of~$\mathfrak g$ (i.e., \emph{Casimir operators}) and polynomial invariants of~${\rm Ad}^*_G$, assumed to be globally defined on~$\mathfrak g^*$. This linear bijection is given by the symmetrization map~$S(\mathfrak g)\to U(\mathfrak g)$, which restricts to the invariant subspace~$S(\mathfrak g)^{\mathfrak g}$; see~\cite{Dixmier} and \cite{Abellanas-Martinez,Campoamor-Rausch}.

For rational or more general invariants, the interpretation as generalized Casimir operators requires additional care~\cite{Abellanas-Martinez,Boyko2006}. In the present paper only polynomial invariants are symmetrized. The set of invariants of~$\mathfrak g$ is denoted by~${\mathop{\rm Inv}\nolimits}(\mathfrak g)$.

A collection of~$N_\mathfrak g$ generically functionally independent invariants $F^l(x_1,\ldots,x_n)$, $l=1,\ldots,N_\mathfrak g$, of~${\rm Ad}^*_G$ is called a \emph{functional basis} of this group, in the local sense on a regular open subset. The collection of symmetrized expressions~$\mathop{\rm Sym}\nolimits F^l(e_1,\ldots,e_n)$, $l=1,\ldots,N_\mathfrak g$, is called a basis of invariants of the Lie algebra~$\mathfrak g$. Thus, we first find invariants of~${\rm Ad}^*_G$ and then transform them into invariants of~$\mathfrak g$ by symmetrization.

Within the classical approach~\cite{Abellanas-Martinez,Linton-Holderbaum,patera-sharp,Pecina-Cruz}, invariants of~${\rm Ad}^*_G$ are locally obtained as first integrals of the coadjoint orbit foliation by solving
\begin{equation}\label{SystemForLieAlgebraInvariants}
X_iF=0,\quad i=1,\ldots,n,
\end{equation}
where~$X_i=c_{ij}^kx_k\partial_{x_j}$. The infinitesimal generator of~$\{\operatorname{Ad}^*_{\exp(\varepsilon e_i)}\}$ is~$-X_i$. On the whole dual space, the system characterizes invariants of the connected group~$G$.

The \emph{algebraic algorithm} for finding invariants of a Lie algebra~$\mathfrak g$ consists of the following four steps (see~\cite{Boyko&Patera&Popovych2009} for details and the examples below):
\begin{list}{\arabic{enumi}.}{\usecounter{enumi}
\labelwidth=3ex\labelsep=1ex\leftmargin=3ex
\topsep1mm\parsep1mm\itemsep=-1mm\partopsep0mm}

\item \emph{Construct a generic matrix~$B(\theta)$ of the group of inner automorphisms of~$\mathfrak g$.}

\item \emph{Write down the fundamental lifted invariant.}

\item \emph{Eliminate the group parameters by normalization.} For the normalization procedure, see~\cite{Fels-Olver,Fels-Olver2} and Proposition~1 of~\cite{Boyko&Patera&Popovych2009}.

\item \emph{Symmetrize the resulting invariants.}
\end{list}

Previous results on the computation of invariants illustrate the effectiveness of this version of the algebraic method~\cite{Boyko2006,Boyko2007a,Boyko&Patera&Popovych2009}. In particular, the algorithm determines the number of basis invariants in the course of their construction.
To improve its efficiency and considerably simplify computations, different coordinate systems on groups of inner automorphisms may be used (canonical coordinates of the first or second kind or special coordinates).

In many cases, the normalization procedure can be substantially simplified by a suitable choice of a basis of lifted invariants and by using various techniques for parameter elimination (empirical techniques, techniques involving additional combinations of lifted invariants, the use of a floating system of normalization equations, etc.).

The fundamental lifted invariant is the row~$\mathcal I=\check x\cdot B(\theta)$, where $B(\theta)$ represents~$\operatorname{Ad}_{g^{-1}}$ for the corresponding element~$g$. As~$g$ varies in~$G$, these matrices range over the group of inner automorphisms.

\section[Invariants of seven-dimensional nilpotent Lie algebras]{Invariants of seven-dimensional nilpotent\\ Lie algebras}\label{inv7al}

We use Seeley's classification as corrected by Gong~\cite{gong1998,seeley1993}. Both the classification and the identifications of parameter values are up to complex Lie algebra isomorphism. Distinct real forms of a complex Lie algebra need not be isomorphic over~$\mathbb R$. For the classification over~$\mathbb R$, as well as an independent classification over~$\mathbb C$, see Gong~\cite{gong1998}. For another treatment of the complex classification by adjoining a derivation to six-dimensional nilpotent Lie algebras, see Magnin~\cite{Magnin2010}. Gong~\cite[Appendix~C]{gong1998} also discusses the incompleteness of the earlier real classification by Romdhani~\cite{Romdhani}. In the complex case, Gong lists 119 isolated indecomposable isomorphism classes and six one-parameter families~\cite[pp.~51--52]{gong1998}.

Both the form of the invariants and the complexity of their computation depend substantially on the choice of a basis of the Lie algebra. A suitable choice of basis can considerably simplify both the invariants themselves and their construction. For a complete flag~$K=(A_i)_{i=0}^n$ with~$A_0=0$, $\dim A_i=i$ and~$A_i$ an ideal of~$A_{i+1}$ for~$i<n$, a basis satisfying~$A_i=\langle e_1,\ldots,e_i\rangle$ is called~$K$-canonical~\cite{Boyko2006,mubarakzyanov1963.1}. For example, Boyko, Patera and Popovych~\cite{Boyko2006} adapted the bases in the classifications of six-dimensional nilpotent Lie algebras~\cite{morozov1958} and six-dimensional solvable Lie algebras with four-dimensional nilradicals~\cite{turkowski1990}.

An important property of~$K$-canonical bases is that all bases corresponding to the same flag are related by triangular linear transformations.

For each nilpotent Lie algebra in Table~\ref{tab:casimir}, we use a~$K$-canonical basis adapted to a flag satisfying~$[\mathfrak g,A_j]\subseteq A_{j-1}$. Such flags can be obtained by refining the upper central series. In these bases the matrices of inner automorphisms are upper unitriangular, which enables successive elimination of the group parameters.

For the nilpotent Lie algebras considered here, polynomial functional bases exist~\cite{Abellanas-Martinez,Ndogmo,Ooms2012}. We obtain polynomial candidates by taking the numerators and denominators of the reduced rational invariants from the successive normalization below.

Below we describe the successive normalization procedure and illustrate it with examples. The results are summarized in Table~\ref{tab:casimir}.

\subsection[Successive normalization in K-canonical bases]{Successive normalization in~$\boldsymbol{K}$-canonical bases}
In the chosen bases, the flag condition can be written as $[e_i,e_j]\in\langle e_1,\ldots,e_{\min(i,j)-1}\rangle$. Let~$A(v)$ be the matrix of~$\operatorname{ad}_v$, with the coordinates of~$[v,e_j]$ in column~$j$. The flag condition makes every~$A(v)$ strictly upper triangular. We can normalize the components of the fundamental lifted invariant successively, without first expanding the entire matrix of inner automorphisms. For an earlier construction of invariants along a complete flag of ideals, see~\cite[Lemma~3]{Dixmier1957} and~\cite[Section~3]{Ooms2012}.

Start with~$y=\check x$. At step~$j$, let
\[
 H_{j-1}(y)=\{v\in\mathfrak g\mid (yA(v))_i=0,\ i<j\}.
\]
If~$(yA(v))_j=0$ for all~$v\in H_{j-1}(y)$, retain the current value of the~$j$th coordinate. Otherwise, choose ${v\in H_{j-1}(y)}$ with~$a=(yA(v))_j\ne0$ on a fixed nonempty Zariski-open set where the selected pivots are nonzero, and normalize the~$j$th coordinate by the finite inner automorphism
\begin{equation*}
 y\longmapsto y\exp(tA(v)),\quad t=-y_j/a.
\end{equation*}
For~$v\in H_{j-1}(y)$, the first~$j-1$ entries of~$yA(v)$ vanish. Strict upper triangularity then gives~$(yA(v)^m)_i=0$ for~$i\le j$ and~$m\ge2$. Hence this transformation preserves the preceding coordinates and sets the~$j$th coordinate to zero. The direction~$v$ is obtained by elementary row operations on the first~$j$ columns of the matrix~$(c_{ik}^l y_l)$, not by solving the differential equations~\eqref{SystemForLieAlgebraInvariants}. The exponential is a finite sum, since~$A(v)^7=0$.

The resulting normalized coordinates are rational functions on a regular open subset of~$\mathfrak g^*$. To see that they are invariants, note that each element of the unipotent group of inner automorphisms is the exponential of an inner derivation. Its action fixes the first~$j-1$ coordinates of~$y$ if and only if the corresponding element~$v$ belongs to~$H_{j-1}(y)$: on the truncated space this follows from~$\exp A-I=A(I+A/2!+\cdots)$, with the second factor invertible. It follows inductively that the normalized values of the retained coordinates are independent of the choice of transformations to the cross-section. The number of retained coordinates equals the codimension of a generic coadjoint orbit.

The numerators and denominators of the reduced rational invariants are polynomial invariants: the relatively prime numerator and denominator can only acquire a common scalar multiplier under a group transformation, and the unipotent group has no nontrivial algebraic characters. A polynomial functional basis is selected from these polynomials by testing functional independence. For a compact presentation, we also use polynomial combinations of normalized coordinates whose expansions are polynomials in the original coordinates. These are obtained by invariantizing polynomial functions on the cross-section.

\subsection{Illustrative examples}\label{illustrative-example2}
In Seeley's classification~\cite[pp.~482--494]{seeley1993}, $2\oplus 1,3,5$ denotes the direct sum of the two-dimensional Abelian Lie algebra and the unique Lie algebra whose upper central series has terms of dimensions~$1$, $3$ and~$5$. After passing to~$K$-canonical bases, we number the non-Abelian entries consecutively and label the Abelian algebra~$\mathfrak n_{7.0}$. The correspondence with Seeley's labels is given in Table~\ref{tab:casimir}.
In the examples below, $A_j=A(e_j)$ with the column convention above. We use canonical coordinates of the second kind,
\[
 B(\theta)=\prod_{j=1}^7\exp(\theta_j A_j),\quad
 \mathcal I=\check x B(\theta),
\]
where the factors are ordered by increasing~$j$ and those corresponding to central basis elements are omitted. The commutation relations and the symmetrized bases of invariants are listed in Table~\ref{tab:casimir}. All normalizations are performed on the indicated open subsets of~$\mathfrak g^*$.

\begin{Example}
For the Lie algebra~$\mathfrak n_{7.2}$, all products~$A_iA_j$ vanish, and hence
\[
 B(\theta)=I_7+\theta_5(E_{36}+E_{47})-\theta_6E_{35}-\theta_7E_{45},
\]
where~$I_7$ is the identity matrix and~$E_{ij}$ denotes the matrix unit with the nonzero entry in position~$(i,j)$.
Hence~$\mathcal I_i=x_i$, $i=1,\ldots,4$, and
\[
 \mathcal I_5=x_5-\theta_6x_3-\theta_7x_4,\quad
 \mathcal I_6=x_6+\theta_5x_3,\quad
 \mathcal I_7=x_7+\theta_5x_4.
\]
On the set~$x_3\ne0$, set~$\theta_7=0$. Then~$\mathcal I_5=\mathcal I_6=0$ gives~$\theta_6=x_5/x_3$ and~$\theta_5=-x_6/x_3$. Substituting~$\theta_5=-x_6/x_3$ into~$\mathcal I_7$ and multiplying the result by~$\mathcal I_3=x_3$, we obtain the polynomial invariant
\[
 F=\mathcal I_3\mathcal I_7-\mathcal I_4\mathcal I_6=x_3x_7-x_4x_6.
\]
Since the coadjoint representation has rank~2, $x_1$, $x_2$, $x_3$, $x_4$ and~$F$ form a functional basis of invariants; their independence follows from~$\partial F/\partial x_7=x_3$.
\end{Example}

\begin{Example}
For the Lie algebra~$\mathfrak n_{7.22}$, we have~$\mathcal I_i=x_i$, $i=1,2,3$, and
\begin{align*}
 \mathcal I_4&=x_4-\theta_6x_3,\quad
 \mathcal I_5=x_5-\theta_6x_4+\tfrac12\theta_6^2x_3,\\
 \mathcal I_7&=x_7+\theta_6x_5-\tfrac12\theta_6^2x_4+\tfrac16\theta_6^3x_3.
\end{align*}
On the set~$x_3\ne0$, the first normalization equation~$\mathcal I_4=0$ gives~$\theta_6=x_4/x_3$. Substitution into~$\mathcal I_5$ and~$\mathcal I_7$ and multiplication by~$2x_3$ and~$3x_3^2$, respectively, yield two polynomial invariants,
\[
 F_1=2x_3x_5-x_4^2,\quad
 F_2=3x_3^2x_7+3x_3x_4x_5-x_4^3.
\]
With~$\theta_5=\theta_7=0$, one has~$\mathcal I_6=x_6+\theta_4x_3$. The second normalization equation~$\mathcal I_6=0$ gives~$\theta_4=-x_6/x_3$ and leaves~$\mathcal I_5$ and~$\mathcal I_7$ unchanged. The coadjoint representation has rank~2, and~$x_1$, $x_2$, $x_3$, $F_1$ and~$F_2$ form a functional basis of invariants, since the Jacobian determinant of~$(F_1,F_2)$ with respect to~$(x_5,x_7)$ equals~$6x_3^3$.
\end{Example}

\begin{Example}
Consider the Lie algebra~$\mathfrak n_{7.81}$, labeled~$2,3,5,7_D$ by Seeley. We set~$\theta_7=0$ and eliminate the four parameters~$\theta_3,\ldots,\theta_6$ by normalization. For this choice, $\mathcal I_i=x_i$, $i=1,2$, and
\begin{align*}
 \mathcal I_3&=x_3-\theta_6x_2,\quad
 \mathcal I_4=x_4+\theta_5x_2,\\
 \mathcal I_5&=x_5-\theta_4x_2-\theta_6x_3+\tfrac12\theta_6^2x_2,\quad
 \mathcal I_6=x_6+\theta_3x_2+\theta_5x_3,\\
 \mathcal I_7&=x_7+\theta_4(x_1+\mathcal I_3)+\theta_5x_1
 +\theta_6x_5-\tfrac12\theta_6^2x_3+\tfrac16\theta_6^3x_2.
\end{align*}
On the set~$x_2\ne0$, we solve the triangular system of normalization equations~$\mathcal I_3=\mathcal I_4=\mathcal I_5=\mathcal I_6=0$ successively,
\begin{align*}
 \theta_6&=\frac{x_3}{x_2},\quad
 \theta_5=-\frac{x_4}{x_2},\quad
 \theta_4=\frac{x_5}{x_2}-\frac{x_3^2}{2x_2^2},\quad
 \theta_3=-\frac{x_6+\theta_5x_3}{x_2}.
\end{align*}
The normalized value of~$\mathcal I_7$ multiplied by~$-6x_2^2$ gives
\[
 F=6x_1x_2x_4-6x_1x_2x_5+3x_1x_3^2-6x_2^2x_7-6x_2x_3x_5+2x_3^3.
\]
The rank of the coadjoint representation is~4 (the principal Pfaffian of~$(c_{ij}^kx_k)$ on indices~$3,4,5,6$ is~$-x_2^2$, and the center is two-dimensional). Thus the four normalization equations are sufficient. Direct substitution into~\eqref{SystemForLieAlgebraInvariants} confirms the invariance of~$F$ under the entire group. Since~$\partial F/\partial x_7=-6x_2^2$, $x_1$, $x_2$ and~$F$ constitute a functional basis of invariants.
\end{Example}

\begin{Example}
Consider the one-parameter family~$\mathfrak n_{7.119}(\xi)$, labeled~$1,3,5,7_M$ by Seeley, where~$\xi\notin\{0,\tfrac12\}$. This example illustrates how the normalization depends on the parameter of a family of Lie algebras. On~$x_1\ne0$, the first two normalization steps use the directions~$e_6$ and~$e_7$ with transformation parameters~$t=x_2/(\xi x_1)$ and~$t=x_3/x_1$, respectively, setting the second and third coordinates to zero. After these two steps the fourth and fifth coordinates are~$x_4-x_2x_3/(\xi x_1)$ and~$x_5-x_3^2/(2x_1)$, respectively. The principal Pfaffian on indices~$2,\ldots,7$ is~$\xi(\xi-1)x_1^3$, while that on indices~$2,3,6,7$ is~$-\xi x_1^2$. This gives generic rank six for admissible~$\xi\ne1$ and rank four for~$\xi=1$. If~$\xi\ne1$, the remaining four noncentral coordinates can be normalized to zero, so the normalized row is~$(x_1,0,0,0,0,0,0)$ and~$x_1$ constitutes a~functional basis of invariants.

If~$\xi=1$, only the sixth and seventh coordinates can be further normalized. The normalized row is
\[
 \left(x_1,0,0,x_4-\frac{x_2x_3}{x_1},x_5-\frac{x_3^2}{2x_1},0,0\right).
\]
Multiplying its fourth and fifth components by~$-x_1$ and~$-2x_1$, respectively, yields the polynomial invariants
\[
 F_1=x_2x_3-x_1x_4,\quad F_2=x_3^2-2x_1x_5.
\]
Thus, at~$\xi=1$, the functional basis consists of~$x_1$, $F_1$ and~$F_2$. Their independence follows from the nonzero Jacobian minor~$2x_1^2$ with respect to~$(x_1,x_4,x_5)$. The number of independent invariants increases from one to three at this exceptional parameter value.

\end{Example}

\subsection[Bases of invariants of seven-dimensional nilpotent Lie algebras]{Bases of invariants of seven-dimensional nilpotent\\ Lie algebras}

\paragraph{Corrections to Seeley's classification over~$\mathbb C$.}
Gong~\cite[p.~3]{gong1998} pointed out four corrections to the original classification over~$\mathbb C$~\cite{seeley1993} (in Seeley's notation):
\begin{enumerate}
\item In~$1,3,7_B$, the relation~$[b,c]=g$ is to be replaced by~$[b,d]=g$; otherwise the entry is isomorphic to~$1,3,7_A$.
\item The entry~$1,3,4,5,7_H$ must be removed because the Jacobi identity for~$(a,b,c)$ fails.
\item In~$1,4,7_E$, the parameter must satisfy~$\xi\ne 0,1$; for either excluded value the center is two-dimensional and the algebra is isomorphic to~$2,4,7_P$.
\item In~$1,3,5,7_S$, the value~$\xi=1$ must be excluded; at this value the center is two-dimensional and the algebra is isomorphic to~$2,3,5,7_D$.
\end{enumerate}
We implement these corrections: $1,3,7_B$ is represented by~$\mathfrak n_{7.104}$ and~$1,3,4,5,7_H$ is omitted. The exceptional classes~$2,4,7_P$ and~$2,3,5,7_D$ are represented by~$\mathfrak n_{7.58}$ and~$\mathfrak n_{7.81}$, respectively. Gong~\cite[pp.~51--52]{gong1998} incorporates the four isolated entries~$\mathfrak n_{7.98}$, $\mathfrak n_{7.117}$, $\mathfrak n_{7.146}$ and~$\mathfrak n_{7.157}$ into the corresponding families, at~$\xi=\tfrac12$, $\xi=\tfrac12$, $\xi=0$ and~$\xi=1$, respectively. The specialization~$\mathfrak n_{7.119}(0)\simeq\mathfrak n_{7.79}$ has a different upper central series and remains a separate entry in both lists. We retain all four isolated entries and exclude their parameter values from the families. Thus Seeley's classification as corrected by Gong has 31 decomposable entries (including the Abelian algebra), 123 isolated indecomposable entries and six one-parameter families.

The resulting parameter domains are (cf.~\cite[pp.~490--494]{seeley1993}):
\begin{gather*}
\mathfrak n_{7.100}(\xi):\quad \xi\notin\{0,1,2,-1,\tfrac12\},\quad
\mathfrak n_{7.119}(\xi):\quad \xi\notin\{0,\tfrac12\},\\
\mathfrak n_{7.120}(\xi):\quad \xi\in\mathbb C,\quad
\mathfrak n_{7.125}(\xi):\quad \xi\ne1,\\
\mathfrak n_{7.147}(\xi):\quad \xi\ne0,\quad
\mathfrak n_{7.159}(\xi):\quad \xi\ne1.
\end{gather*}
For~$\mathfrak n_{7.100}$, the equivalent values of~$\xi$ are
\[
 \xi,\quad \xi^{-1},\quad \frac{\xi}{\xi-1},\quad \frac{\xi-1}{\xi},\quad 1-\xi,\quad \frac{1}{1-\xi}.
\]
For~$\mathfrak n_{7.147}$, the parameter values~$\xi$ and~$\xi^{-1}$ are equivalent. In the other four families distinct admissible parameter values give nonisomorphic algebras. The excluded values~$2,-1,\tfrac12$ for~$\mathfrak n_{7.100}$ represent~$\mathfrak n_{7.98}$; the excluded value~$\tfrac12$ for~$\mathfrak n_{7.119}$ represents~$\mathfrak n_{7.117}$, and its excluded value~$0$ represents~$\mathfrak n_{7.79}$. The values~$0$ for~$\mathfrak n_{7.147}$ and~$1$ for~$\mathfrak n_{7.159}$ correspond to~$\mathfrak n_{7.146}$ and~$\mathfrak n_{7.157}$, respectively. The two remaining excluded specializations are accounted for by the isomorphisms~$\mathfrak n_{7.100}(0)\simeq\mathfrak n_{7.58}\simeq\mathfrak n_{7.100}(1)$ and~$\mathfrak n_{7.125}(1)\simeq\mathfrak n_{7.81}$.

Table~\ref{tab:casimir} comprises 160 entries, including the Abelian algebra and six one-parameter families. The change from Seeley's basis to the tabulated basis is given in the remark following the table. Each label in the first column is accompanied by the corresponding designation in Seeley's list. All brackets not listed are zero, up to skew-symmetry. Each listed set is a polynomial functional basis, not necessarily a generating set of the algebra of polynomial invariants. For a description of the latter algebras in the indecomposable cases, see~\cite{Ooms2009,Ooms2012}. For example, for~$\mathfrak n_{7.9}$ let~$p_{ij}=x_i x_{j+3}-x_j x_{i+3}$ for~$1\le i<j\le3$. The polynomial~$p_{23}$ is invariant and satisfies $x_1p_{23}=x_2p_{13}-x_3p_{12}$. At~$(0,1,0,0,0,0,0)$ and~$(0,1,0,0,0,1,0)$ all five tabulated invariants agree, whereas~$p_{23}$ takes the values~$0$ and~$1$. Thus~$p_{23}$ lies in the field generated by the tabulated invariants and outside their polynomial algebra. See the entry for~$g_{7,4.2}$ in~\cite{Ooms2012}. In decomposable cases, the polynomial invariants of a direct sum are obtained by taking tensor products of the polynomial invariant algebras of its summands.

For every entry, we construct the normalized coordinates by the successive normalization described above. We first obtain a polynomial functional basis from their numerators and denominators and then use invariantization of suitable polynomials on the cross-section to retain compact expressions in the table. The equations~\eqref{SystemForLieAlgebraInvariants} are used only to verify the resulting polynomial invariants. For the parameter families, we split the parameter domain whenever a coefficient used in the normalization vanishes identically or a normalization formula ceases to be defined. In addition to the generic computations for the six families, we perform separate normalizations at the admissible values~$\xi=1$ for~$\mathfrak n_{7.119}$ and~$\mathfrak n_{7.147}$ and~$\xi=0$ for~$\mathfrak n_{7.120}$, $\mathfrak n_{7.125}$ and~$\mathfrak n_{7.159}$. The number of retained coordinates increases from one to three in four of these cases. For~$\mathfrak n_{7.125}(0)$ the generic normalization uses a pivot proportional to~$\xi$, but the principal Pfaffian on indices~$2,\ldots,7$ is~$(\xi-1)x_1^3$. Hence the pivot must be replaced at~$\xi=0$, where the index is still one and~$x_1$ remains a functional basis. The table includes the resulting bases for all admissible parameter values. The accompanying computational data record the normalizing transformations and the normalized coordinates for each case.

For each entry, we verified the Jacobi identity and strict triangularity. For every listed polynomial~$F$ in the coordinates~$x_i$, we checked that $C(x)\nabla F=0$. We computed the generic rank of~$C(x)$ from its principal Pfaffians and verified functional independence by an exact nonzero Jacobian minor of order~$7-r$, where~$r=\operatorname{rank}C(x)$ is the generic rank.

The displayed polynomials in the~$e_i$ are understood as their symmetrizations in the universal enveloping algebra.

\begingroup\footnotesize
\setlength{\tabcolsep}{2.5pt}
\renewcommand{\arraystretch}{1.05}
\setlength{\LTleft}{\fill}\setlength{\LTright}{\fill}
\begin{longtable}{|>{\raggedright\arraybackslash\leavevmode}p{25mm}|>{\raggedright\arraybackslash\leavevmode}p{65mm}|>{\raggedright\arraybackslash\leavevmode}p{64mm}|}
\caption{Casimir operators for seven-dimensional complex nilpotent Lie algebras}\label{tab:casimir}\\
\hline Lie algebra & Nonzero commutation relations & Casimir operators \\\hline
\endfirsthead
\hline Lie algebra & Nonzero commutation relations & Casimir operators \\\hline
\endhead
$\mathfrak n_{7.0}$\newline{\scriptsize $7$} & None & $e_1$, $e_2$, $e_3$, $e_4$, $e_5$, $e_6$, $e_7$ \\\hline
$\mathfrak{n}_{7.1}$\newline {\scriptsize $4 \oplus 1,3 $} & \mbox{$[e_{6},e_{7}]=e_{5}$} & $e_{1}$, $e_{2}$, $e_{3}$, $e_{4}$, $e_{5}$\\\hline
$\mathfrak{n}_{7.2}$\newline {\scriptsize $2\oplus 2,5$} & \mbox{$[e_{5},e_{6}]=e_{3}$}, \mbox{$[e_{5},e_{7}]=e_{4}$} & $e_{1}$, $e_{2}$, $e_{3}$, $e_{4}$, $e_{3} e_{7}- e_{4} e_{6}$\\\hline
$\mathfrak{n}_{7.3}$\newline {\scriptsize $1\oplus 3,6$} & \mbox{$[e_{5},e_{6}]=e_{2}$}, \mbox{$[e_{5},e_{7}]=e_{3}$}, \mbox{$[e_{6},e_{7}]=e_{4}$} & $e_{1}$, $e_{2}$, $e_{3}$, $e_{4}$, $e_{2} e_{7}- e_{3} e_{6}+e_{4} e_{5}$\\\hline
$\mathfrak{n}_{7.4}$\newline {\scriptsize $3\oplus 1,2,4$} & \mbox{$[e_{5},e_{7}]=e_{4}$}, \mbox{$[e_{6},e_{7}]=e_{5}$} & $e_{1}$, $e_{2}$, $e_{3}$, $e_{4}$, $- 2 e_{4} e_{6}+e_{5}^{2}$\\\hline
$\mathfrak{n}_{7.5}$\newline {\scriptsize $2\oplus 2,3,5$} & \mbox{$[e_{5},e_{6}]=e_{3}$}, \mbox{$[e_{5},e_{7}]=e_{4}$}, \mbox{$[e_{6},e_{7}]=e_{5}$} & $e_{1}$, $e_{2}$, $e_{3}$, $e_{4}$, $2 e_{3} e_{7}- 2 e_{4} e_{6}+e_{5}^{2}$\\\hline
$\mathfrak{n}_{7.6}$\newline {\scriptsize $2 \oplus 1,5 $} & \mbox{$[e_{4},e_{5}]=e_{3}$}, \mbox{$[e_{6},e_{7}]=e_{3}$} & $e_{1}$, $e_{2}$, $e_{3}$\\\hline
$\mathfrak{n}_{7.7}$\newline {\scriptsize $1 \oplus 2,6 $} & \mbox{$[e_{4},e_{5}]=e_{3}$}, \mbox{$[e_{5},e_{6}]=e_{2}$}, \mbox{$[e_{6},e_{7}]=e_{3}$} & $e_{1}$, $e_{2}$, $e_{3}$\\\hline
$\mathfrak{n}_{7.8}$\newline {\scriptsize $1\oplus 1,3\oplus 1,3$} & \mbox{$[e_{4},e_{5}]=e_{2}$}, \mbox{$[e_{6},e_{7}]=e_{3}$} & $e_{1}$, $e_{2}$, $e_{3}$\\\hline
$\mathfrak{n}_{7.9}$\newline {\scriptsize $3,7_{A}$} & \mbox{$[e_{4},e_{7}]=e_{1}$}, \mbox{$[e_{5},e_{7}]=e_{2}$}, \mbox{$[e_{6},e_{7}]=e_{3}$} & $e_{1}$, $e_{2}$, $e_{3}$, $e_{1} e_{5}- e_{2} e_{4}$, $e_{1} e_{6}- e_{3} e_{4}$\\\hline
$\mathfrak{n}_{7.10}$\newline {\scriptsize $3,7_{B}$} & \mbox{$[e_{4},e_{5}]=e_{1}$}, \mbox{$[e_{5},e_{7}]=e_{2}$}, \mbox{$[e_{6},e_{7}]=e_{3}$} & $e_{1}$, $e_{2}$, $e_{3}$\\\hline
$\mathfrak{n}_{7.11}$\newline {\scriptsize $3,7_{C}$} & \mbox{$[e_{4},e_{5}]=e_{3}$}, \mbox{$[e_{4},e_{7}]=e_{1}$}, \mbox{$[e_{5},e_{7}]=e_{2}$}, \mbox{$[e_{6},e_{7}]=e_{3}$} & $e_{1}$, $e_{2}$, $e_{3}$\\\hline
$\mathfrak{n}_{7.12}$\newline {\scriptsize $3,7_{D}$} & \mbox{$[e_{4},e_{5}]=e_{3}$}, \mbox{$[e_{4},e_{6}]=e_{1}$}, \mbox{$[e_{5},e_{7}]=e_{2}$}, \mbox{$[e_{6},e_{7}]=e_{3}$} & $e_{1}$, $e_{2}$, $e_{3}$\\\hline
$\mathfrak{n}_{7.13}$\newline {\scriptsize $2 \oplus 1, 3, 5$} & \mbox{$[e_{4},e_{7}]=e_{3}$}, \mbox{$[e_{5},e_{6}]=e_{3}$}, \mbox{$[e_{6},e_{7}]=e_{4}$} & $e_{1}$, $e_{2}$, $e_{3}$\\\hline
$\mathfrak{n}_{7.14}$\newline {\scriptsize $1 \oplus 2, 4, 6_A$} & \mbox{$[e_{4},e_{6}]=e_{2}$}, \mbox{$[e_{5},e_{6}]=e_{3}$}, \mbox{$[e_{6},e_{7}]=e_{5}$} & $e_{1}$, $e_{2}$, $e_{3}$, $- e_{2} e_{5}+e_{3} e_{4}$, $2 e_{3} e_{7}+e_{5}^{2}$\\\hline
$\mathfrak{n}_{7.15}$\newline {\scriptsize $1 \oplus 2, 4, 6_B$} & \mbox{$[e_{4},e_{7}]=e_{2}$}, \mbox{$[e_{5},e_{6}]=e_{3}$}, \mbox{$[e_{6},e_{7}]=e_{5}$} & $e_{1}$, $e_{2}$, $e_{3}$\\\hline
$\mathfrak{n}_{7.16}$\newline {\scriptsize $1 \oplus 2, 4, 6_C$} & \mbox{$[e_{4},e_{6}]=e_{2}$}, \mbox{$[e_{4},e_{7}]=e_{3}$}, \mbox{$[e_{5},e_{6}]=e_{3}$}, \mbox{$[e_{6},e_{7}]=e_{5}$} & $e_{1}$, $e_{2}$, $e_{3}$\\\hline
$\mathfrak{n}_{7.17}$\newline {\scriptsize $1 \oplus 2, 4, 6_D$} & \mbox{$[e_{4},e_{6}]=e_{2}$}, \mbox{$[e_{5},e_{6}]=e_{3}$}, \mbox{$[e_{5},e_{7}]=e_{2}$}, \mbox{$[e_{6},e_{7}]=e_{5}$} & $e_{1}$, $e_{2}$, $e_{3}$\\\hline
$\mathfrak{n}_{7.18}$\newline {\scriptsize $1 \oplus 2, 4, 6_E$} & \mbox{$[e_{4},e_{6}]=e_{3}$}, \mbox{$[e_{5},e_{6}]=e_{3}$}, \mbox{$[e_{5},e_{7}]=e_{2}$}, \mbox{$[e_{6},e_{7}]=e_{5}$} & $e_{1}$, $e_{2}$, $e_{3}$\\\hline
$\mathfrak{n}_{7.19}$\newline {\scriptsize $3,5,7_A$} & \mbox{$[e_{4},e_{6}]=e_{2}$}, \mbox{$[e_{4},e_{7}]=e_{1}$}, \mbox{$[e_{5},e_{6}]=e_{3}$}, \mbox{$[e_{6},e_{7}]=e_{5}$} & $e_{1}$, $e_{2}$, $e_{3}$\\\hline
$\mathfrak{n}_{7.20}$\newline {\scriptsize $3,5,7_B$} & \mbox{$[e_{4},e_{6}]=e_{2}$}, \mbox{$[e_{5},e_{6}]=e_{3}$}, \mbox{$[e_{5},e_{7}]=e_{1}$}, \mbox{$[e_{6},e_{7}]=e_{5}$} & $e_{1}$, $e_{2}$, $e_{3}$\\\hline
$\mathfrak{n}_{7.21}$\newline {\scriptsize $3,5,7_C$} & \mbox{$[e_{4},e_{6}]=e_{2}$}, \mbox{$[e_{4},e_{7}]=e_{3}$}, \mbox{$[e_{5},e_{6}]=e_{3}$}, \mbox{$[e_{5},e_{7}]=e_{1}$}, \mbox{$[e_{6},e_{7}]=e_{5}$} & $e_{1}$, $e_{2}$, $e_{3}$\\\hline
$\mathfrak{n}_{7.22}$\newline {\scriptsize $2 \oplus 1, 2, 3, 5_A$} & \mbox{$[e_{4},e_{6}]=e_{3}$}, \mbox{$[e_{5},e_{6}]=e_{4}$}, \mbox{$[e_{6},e_{7}]=e_{5}$} & $e_{1}$, $e_{2}$, $e_{3}$, $2 e_{3} e_{5}- e_{4}^{2}$, $3 e_{3}^{2} e_{7}+3 e_{3} e_{4} e_{5}- e_{4}^{3}$\\\hline
$\mathfrak{n}_{7.23}$\newline {\scriptsize $2 \oplus 1, 2, 3, 5_B$} & \mbox{$[e_{4},e_{6}]=e_{3}$}, \mbox{$[e_{5},e_{6}]=e_{4}$}, \mbox{$[e_{5},e_{7}]=e_{3}$}, \mbox{$[e_{6},e_{7}]=e_{5}$} & $e_{1}$, $e_{2}$, $e_{3}$\\\hline
$\mathfrak{n}_{7.24}$\newline {\scriptsize $1\oplus 2,3,4,6$} & \mbox{$[e_{4},e_{6}]=e_{3}$}, \mbox{$[e_{5},e_{6}]=e_{4}$}, \mbox{$[e_{5},e_{7}]=e_{2}$}, \mbox{$[e_{6},e_{7}]=e_{5}$} & $e_{1}$, $e_{2}$, $e_{3}$\\\hline
$\mathfrak{n}_{7.25}$\newline {\scriptsize $2, 7_A$} & \mbox{$[e_{3},e_{4}]=e_{1}$}, \mbox{$[e_{5},e_{6}]=e_{1}$}, \mbox{$[e_{5},e_{7}]=e_{2}$} & $e_{1}$, $e_{2}$, $- e_{1} e_{7}+e_{2} e_{6}$\\\hline
$\mathfrak{n}_{7.26}$\newline {\scriptsize $2, 7_B$} & \mbox{$[e_{3},e_{5}]=e_{1} + e_{2}$}, \mbox{$[e_{3},e_{7}]=e_{1}$}, \mbox{$[e_{4},e_{5}]=e_{1} + e_{2}$}, \mbox{$[e_{4},e_{6}]=e_{2}$} & $e_{1}$, $e_{2}$, $e_1^2e_6-e_1e_2e_5+e_1e_2e_6+e_1e_2e_7+e_2^2e_7$\\\hline
$\mathfrak{n}_{7.27}$\newline {\scriptsize $1 \oplus 1, 4, 6$} & \mbox{$[e_{3},e_{4}]=e_{2}$}, \mbox{$[e_{5},e_{6}]=e_{2}$}, \mbox{$[e_{6},e_{7}]=e_{5}$} & $e_{1}$, $e_{2}$, $2 e_{2} e_{7}+e_{5}^{2}$\\\hline
$\mathfrak{n}_{7.28}$\newline {\scriptsize $1,2,4 \oplus 1, 3$} & \mbox{$[e_{3},e_{4}]=e_{1}$}, \mbox{$[e_{5},e_{6}]=e_{2}$}, \mbox{$[e_{6},e_{7}]=e_{5}$} & $e_{1}$, $e_{2}$, $2 e_{2} e_{7}+e_{5}^{2}$\\\hline
$\mathfrak{n}_{7.29}$\newline {\scriptsize $2,5, 7_A$} & \mbox{$[e_{3},e_{6}]=e_{1}$}, \mbox{$[e_{4},e_{7}]=e_{2}$}, \mbox{$[e_{5},e_{6}]=e_{2}$}, \mbox{$[e_{6},e_{7}]=e_{5}$} & $e_{1}$, $e_{2}$, $e_{1} e_{5}- e_{2} e_{3}$\\\hline
$\mathfrak{n}_{7.30}$\newline {\scriptsize $2,5, 7_B$} & \mbox{$[e_{3},e_{6}]=e_{1}$}, \mbox{$[e_{4},e_{7}]=e_{1}$}, \mbox{$[e_{5},e_{6}]=e_{2}$}, \mbox{$[e_{6},e_{7}]=e_{5}$} & $e_{1}$, $e_{2}$, $e_{1} e_{5}- e_{2} e_{3}$\\\hline
$\mathfrak{n}_{7.31}$\newline {\scriptsize $2,5, 7_C$} & \mbox{$[e_{3},e_{7}]=e_{2}$}, \mbox{$[e_{4},e_{7}]=e_{1}$}, \mbox{$[e_{5},e_{6}]=e_{2}$}, \mbox{$[e_{6},e_{7}]=e_{5}$} & $e_{1}$, $e_{2}$, $- e_{1} e_{3}+e_{2} e_{4}$\\\hline
$\mathfrak{n}_{7.32}$\newline {\scriptsize $2,5, 7_D$} & \mbox{$[e_{3},e_{7}]=e_{1}$}, \mbox{$[e_{4},e_{6}]=e_{1}$}, \mbox{$[e_{4},e_{7}]=e_{2}$}, \mbox{$[e_{5},e_{6}]=e_{2}$}, \mbox{$[e_{6},e_{7}]=e_{5}$} & $e_{1}$, $e_{2}$, $- e_{1}^{2} e_{5}+e_{1} e_{2} e_{4}- e_{2}^{2} e_{3}$\\\hline
$\mathfrak{n}_{7.33}$\newline {\scriptsize $2,5, 7_E$} & \mbox{$[e_{3},e_{4}]=e_{2}$}, \mbox{$[e_{4},e_{7}]=e_{1}$}, \mbox{$[e_{5},e_{6}]=e_{2}$}, \mbox{$[e_{6},e_{7}]=e_{5}$} & $e_{1}$, $e_{2}$, $2 e_{1} e_{3}+2 e_{2} e_{7}+e_{5}^{2}$\\\hline
$\mathfrak{n}_{7.34}$\newline {\scriptsize $2,5, 7_F$} & \mbox{$[e_{3},e_{4}]=e_{2}$}, \mbox{$[e_{4},e_{6}]=e_{1}$}, \mbox{$[e_{5},e_{6}]=e_{2}$}, \mbox{$[e_{6},e_{7}]=e_{5}$} & $e_{1}$, $e_{2}$, $2 e_{2} e_{7}+e_{5}^{2}$\\\hline
$\mathfrak{n}_{7.35}$\newline {\scriptsize $2,5, 7_G$} & \mbox{$[e_{3},e_{4}]=e_{2}$}, \mbox{$[e_{3},e_{6}]=e_{1}$}, \mbox{$[e_{4},e_{7}]=e_{1}$}, \mbox{$[e_{5},e_{6}]=e_{2}$}, \mbox{$[e_{6},e_{7}]=e_{5}$} & $e_{1}$, $e_{2}$, $- 2 e_{1}^{2} e_{5}+2 e_{1} e_{2} e_{3}+2 e_{2}^{2} e_{7} +e_{2} e_{5}^{2}$\\\hline
$\mathfrak{n}_{7.36}$\newline {\scriptsize $2,5, 7_H$} & \mbox{$[e_{3},e_{4}]=e_{1}$}, \mbox{$[e_{4},e_{7}]=e_{2}$}, \mbox{$[e_{5},e_{6}]=e_{2}$}, \mbox{$[e_{6},e_{7}]=e_{5}$} & $e_{1}$, $e_{2}$, $2 e_{1} e_{2} e_{7}+e_{1} e_{5}^{2}+2 e_{2}^{2} e_{3}$\\\hline
$\mathfrak{n}_{7.37}$\newline {\scriptsize $2,5, 7_I$} & \mbox{$[e_{3},e_{6}]=e_{1}$}, \mbox{$[e_{4},e_{6}]=e_{2}$}, \mbox{$[e_{5},e_{6}]=e_{2}$}, \mbox{$[e_{5},e_{7}]=e_{1}$}, \mbox{$[e_{6},e_{7}]=e_{5}$} & $e_{1}$, $e_{2}$, $e_{1} e_{4}- e_{2} e_{3}$\\\hline
$\mathfrak{n}_{7.38}$\newline {\scriptsize $2,5, 7_J$} & \mbox{$[e_{3},e_{6}]=e_{1}$}, \mbox{$[e_{4},e_{7}]=e_{2}$}, \mbox{$[e_{5},e_{6}]=e_{2}$}, \mbox{$[e_{5},e_{7}]=e_{1}$}, \mbox{$[e_{6},e_{7}]=e_{5}$} & $e_{1}$, $e_{2}$, $- e_{1}^{2} e_{4}+e_{1} e_{2} e_{5}- e_{2}^{2} e_{3}$\\\hline
$\mathfrak{n}_{7.39}$\newline {\scriptsize $2,5, 7_K$} & \mbox{$[e_{3},e_{4}]=e_{2}$}, \mbox{$[e_{5},e_{6}]=e_{2}$}, \mbox{$[e_{5},e_{7}]=e_{1}$}, \mbox{$[e_{6},e_{7}]=e_{5}$} & $e_{1}$, $e_{2}$, $- 2 e_{1} e_{6}+2 e_{2} e_{7}+e_{5}^{2}$\\\hline
$\mathfrak{n}_{7.40}$\newline {\scriptsize $2,5, 7_L$} & \mbox{$[e_{3},e_{4}]=e_{2}$}, \mbox{$[e_{4},e_{6}]=e_{1}$}, \mbox{$[e_{5},e_{6}]=e_{2}$}, \mbox{$[e_{5},e_{7}]=e_{1}$}, \mbox{$[e_{6},e_{7}]=e_{5}$} & $e_{1}$, $e_{2}$, $- 2 e_{1}^{2} e_{3}- 2 e_{1} e_{2} e_{6}+2 e_{2}^{2} e_{7} +e_{2} e_{5}^{2}$\\\hline
$\mathfrak{n}_{7.41}$\newline {\scriptsize $1 \oplus 1,3,6_A$} & \mbox{$[e_{3},e_{7}]=- e_{2}$}, \mbox{$[e_{4},e_{5}]=e_{2}$}, \mbox{$[e_{5},e_{6}]=e_{3}$}, \mbox{$[e_{6},e_{7}]=e_{4}$} & $e_{1}$, $e_{2}$, $e_{2} e_{6}+e_{3} e_{4}$\\\hline
$\mathfrak{n}_{7.42}$\newline {\scriptsize $1 \oplus 1,3,6_B$} & \mbox{$[e_{3},e_{5}]=- e_{2}$}, \mbox{$[e_{4},e_{6}]=e_{2}$}, \mbox{$[e_{5},e_{6}]=e_{3}$}, \mbox{$[e_{6},e_{7}]=e_{4}$} & $e_{1}$, $e_{2}$, $2 e_{2} e_{7}+e_{4}^{2}$\\\hline
$\mathfrak{n}_{7.43}$\newline {\scriptsize $2,4, 7_A$} & \mbox{$[e_{3},e_{6}]=e_{1}$}, \mbox{$[e_{4},e_{6}]=e_{2}$}, \mbox{$[e_{5},e_{6}]=e_{3}$}, \mbox{$[e_{6},e_{7}]=e_{4}$} & $e_{1}$, $e_{2}$, $e_{1} e_{4}- e_{2} e_{3}$, $2 e_{1} e_{5}- e_{3}^{2}$, $2 e_{1}^{2} e_{7}+2 e_{1} e_{3} e_{4}- e_{2} e_{3}^{2}$\\\hline
$\mathfrak{n}_{7.44}$\newline {\scriptsize $2,4, 7_B$} & \mbox{$[e_{3},e_{5}]=e_{1}$}, \mbox{$[e_{4},e_{6}]=e_{2}$}, \mbox{$[e_{5},e_{6}]=e_{3}$}, \mbox{$[e_{6},e_{7}]=e_{4}$} & $e_{1}$, $e_{2}$, $2 e_{2} e_{7}+e_{4}^{2}$\\\hline
$\mathfrak{n}_{7.45}$\newline {\scriptsize $2,4, 7_C$} & \mbox{$[e_{3},e_{5}]=e_{1}$}, \mbox{$[e_{3},e_{6}]=e_{2}$}, \mbox{$[e_{4},e_{6}]=e_{1}$}, \mbox{$[e_{5},e_{6}]=e_{3}$}, \mbox{$[e_{6},e_{7}]=e_{4}$} & $e_{1}$, $e_{2}$, $2 e_{1} e_{7}+e_{4}^{2}$\\\hline
$\mathfrak{n}_{7.46}$\newline {\scriptsize $2,4, 7_D$} & \mbox{$[e_{3},e_{6}]=e_{1}$}, \mbox{$[e_{3},e_{7}]=- e_{2}$}, \mbox{$[e_{4},e_{5}]=e_{2}$}, \mbox{$[e_{5},e_{6}]=e_{3}$}, \mbox{$[e_{6},e_{7}]=e_{4}$} & $e_{1}$, $e_{2}$, $e_{1} e_{7}+e_{2} e_{6}+e_{3} e_{4}$\\\hline
$\mathfrak{n}_{7.47}$\newline {\scriptsize $2,4, 7_E$} & \mbox{$[e_{3},e_{6}]=e_{1}$}, \mbox{$[e_{3},e_{7}]=e_{2}$}, \mbox{$[e_{4},e_{5}]=- e_{2}$}, \mbox{$[e_{4},e_{6}]=e_{1}$}, \mbox{$[e_{5},e_{6}]=e_{3}$}, \mbox{$[e_{6},e_{7}]=e_{4}$} & $e_{1}$, $e_{2}$, $- e_{1} e_{5}+e_{1} e_{7}- e_{2} e_{6}+e_{3} e_{4}$\\\hline
$\mathfrak{n}_{7.48}$\newline {\scriptsize $2,4, 7_F$} & \mbox{$[e_{3},e_{5}]=- e_{1}$}, \mbox{$[e_{4},e_{7}]=e_{2}$}, \mbox{$[e_{5},e_{6}]=e_{3}$}, \mbox{$[e_{6},e_{7}]=e_{4}$} & $e_{1}$, $e_{2}$, $- 2 e_{1} e_{2} e_{6}+e_{1} e_{4}^{2}+e_{2} e_{3}^{2}$\\\hline
$\mathfrak{n}_{7.49}$\newline {\scriptsize $2,4, 7_G$} & \mbox{$[e_{3},e_{5}]=- e_{1}$}, \mbox{$[e_{4},e_{6}]=e_{1}$}, \mbox{$[e_{4},e_{7}]=e_{2}$}, \mbox{$[e_{5},e_{6}]=e_{3}$}, \mbox{$[e_{6},e_{7}]=e_{4}$} & $e_{1}$, $e_{2}$, $2 e_{1}^{2} e_{7}- 2 e_{1} e_{2} e_{6}+e_{1} e_{4}^{2} +e_{2} e_{3}^{2}$\\\hline
$\mathfrak{n}_{7.50}$\newline {\scriptsize $2,4, 7_H$} & \mbox{$[e_{3},e_{5}]=- e_{1}$}, \mbox{$[e_{3},e_{6}]=- e_{2}$}, \mbox{$[e_{4},e_{6}]=e_{1}$}, \mbox{$[e_{4},e_{7}]=e_{2}$}, \mbox{$[e_{5},e_{6}]=e_{3}$}, \mbox{$[e_{6},e_{7}]=e_{4}$} & $e_{1}$, $e_{2}$, $2 e_{1}^{2} e_{7}- 2 e_{1} e_{2} e_{6}+e_{1} e_{4}^{2} +2 e_{2}^{2} e_{5}+e_{2} e_{3}^{2}$\\\hline
$\mathfrak{n}_{7.51}$\newline {\scriptsize $2,4, 7_I$} & \mbox{$[e_{3},e_{5}]=- e_{1}$}, \mbox{$[e_{3},e_{7}]=- e_{2}$}, \mbox{$[e_{4},e_{5}]=e_{2}$}, \mbox{$[e_{5},e_{6}]=e_{3}$}, \mbox{$[e_{6},e_{7}]=e_{4}$} & $e_{1}$, $e_{2}$, $e_{1} e_{4}^{2}+2 e_{2}^{2} e_{6}+2 e_{2} e_{3} e_{4}$\\\hline
$\mathfrak{n}_{7.52}$\newline {\scriptsize $2,4, 7_J$} & \mbox{$[e_{3},e_{5}]=- e_{1}$}, \mbox{$[e_{3},e_{7}]=- e_{2}$}, \mbox{$[e_{4},e_{5}]=e_{2}$}, \mbox{$[e_{4},e_{6}]=e_{2}$}, \mbox{$[e_{5},e_{6}]=e_{3}$}, \mbox{$[e_{6},e_{7}]=e_{4}$} & $e_{1}$, $e_{2}$, $2 e_{1} e_{2} e_{7}+e_{1} e_{4}^{2}- 2 e_{2}^{2} e_{5} +2 e_{2}^{2} e_{6}+2 e_{2} e_{3} e_{4}$\\\hline
$\mathfrak{n}_{7.53}$\newline {\scriptsize $2,4, 7_K$} & \mbox{$[e_{3},e_{5}]=- e_{1}$}, \mbox{$[e_{3},e_{7}]=- e_{2}$}, \mbox{$[e_{4},e_{5}]=e_{2}$}, \mbox{$[e_{4},e_{6}]=e_{1}$}, \mbox{$[e_{5},e_{6}]=e_{3}$}, \mbox{$[e_{6},e_{7}]=e_{4}$} & $e_{1}$, $e_{2}$, $2 e_{1}^{2} e_{7}- 2 e_{1} e_{2} e_{5}+e_{1} e_{4}^{2} +2 e_{2}^{2} e_{6}+2 e_{2} e_{3} e_{4}$\\\hline
$\mathfrak{n}_{7.54}$\newline {\scriptsize $2,4, 7_L$} & \mbox{$[e_{3},e_{6}]=- e_{1}$}, \mbox{$[e_{4},e_{6}]=e_{2}$}, \mbox{$[e_{5},e_{6}]=e_{3}$}, \mbox{$[e_{5},e_{7}]=e_{2}$}, \mbox{$[e_{6},e_{7}]=e_{4}$} & $e_{1}$, $e_{2}$, $e_{1} e_{4}+e_{2} e_{3}$\\\hline
$\mathfrak{n}_{7.55}$\newline {\scriptsize $2,4, 7_M$} & \mbox{$[e_{3},e_{6}]=- e_{2}$}, \mbox{$[e_{4},e_{7}]=e_{1}$}, \mbox{$[e_{5},e_{6}]=e_{3}$}, \mbox{$[e_{5},e_{7}]=e_{2}$}, \mbox{$[e_{6},e_{7}]=e_{4}$} & $e_{1}$, $e_{2}$, $2 e_{1} e_{2} e_{5}+e_{1} e_{3}^{2}- 2 e_{2}^{2} e_{4}$\\\hline
$\mathfrak{n}_{7.56}$\newline {\scriptsize $2,4, 7_N$} & \mbox{$[e_{3},e_{6}]=- e_{1}$}, \mbox{$[e_{4},e_{7}]=e_{1}$}, \mbox{$[e_{5},e_{6}]=e_{3}$}, \mbox{$[e_{5},e_{7}]=e_{2}$}, \mbox{$[e_{6},e_{7}]=e_{4}$} & $e_{1}$, $e_{2}$, $2 e_{1} e_{5}- 2 e_{2} e_{4}+e_{3}^{2}$\\\hline
$\mathfrak{n}_{7.57}$\newline {\scriptsize $2,4, 7_O$} & \mbox{$[e_{3},e_{6}]=- e_{1}$}, \mbox{$[e_{4},e_{6}]=e_{2}$}, \mbox{$[e_{4},e_{7}]=e_{1}$}, \mbox{$[e_{5},e_{6}]=e_{3}$}, \mbox{$[e_{5},e_{7}]=e_{2}$}, \mbox{$[e_{6},e_{7}]=e_{4}$} & $e_{1}$, $e_{2}$, $- 2 e_{1}^{2} e_{5}+2 e_{1} e_{2} e_{4}- e_{1} e_{3}^{2} +2 e_{2}^{2} e_{3}$\\\hline
$\mathfrak{n}_{7.58}$\newline {\scriptsize $2,4, 7_P$} & \mbox{$[e_{3},e_{7}]=- e_{1}$}, \mbox{$[e_{4},e_{5}]=e_{1}$}, \mbox{$[e_{5},e_{6}]=e_{3}$}, \mbox{$[e_{5},e_{7}]=e_{2}$}, \mbox{$[e_{6},e_{7}]=e_{4}$} & $e_{1}$, $e_{2}$, $e_{1} e_{6}+e_{3} e_{4}$\\\hline
$\mathfrak{n}_{7.59}$\newline {\scriptsize $2,4, 7_Q$} & \mbox{$[e_{3},e_{6}]=- e_{2}$}, \mbox{$[e_{3},e_{7}]=- e_{1}$}, \mbox{$[e_{4},e_{5}]=e_{1}$}, \mbox{$[e_{5},e_{6}]=e_{3}$}, \mbox{$[e_{5},e_{7}]=e_{2}$}, \mbox{$[e_{6},e_{7}]=e_{4}$} & $e_{1}$, $e_{2}$, $- e_{1}^{2} e_{6}+e_{1} e_{2} e_{7}- e_{1} e_{3} e_{4} +e_{2}^{2} e_{4}$\\\hline
$\mathfrak{n}_{7.60}$\newline {\scriptsize $2,4, 7_R$} & \mbox{$[e_{3},e_{6}]=- e_{2}$}, \mbox{$[e_{3},e_{7}]=- e_{1}$}, \mbox{$[e_{4},e_{5}]=e_{1}$}, \mbox{$[e_{4},e_{6}]=e_{2}$}, \mbox{$[e_{5},e_{6}]=e_{3}$}, \mbox{$[e_{5},e_{7}]=e_{2}$}, \mbox{$[e_{6},e_{7}]=e_{4}$} & $e_{1}$, $e_{2}$, $- e_{1}^{2} e_{6}+e_{1} e_{2} e_{5}+e_{1} e_{2} e_{7} - e_{1} e_{3} e_{4}+e_{2}^{2} e_{3}+e_{2}^{2} e_{4}$\\\hline
$\mathfrak{n}_{7.61}$\newline {\scriptsize $1 \oplus 1,3,4,6_A$} & \mbox{$[e_{3},e_{7}]=e_{2}$}, \mbox{$[e_{4},e_{6}]=e_{2}$}, \mbox{$[e_{5},e_{6}]=e_{4}$}, \mbox{$[e_{6},e_{7}]=e_{5}$} & $e_{1}$, $e_{2}$, $2 e_{2} e_{5}- e_{4}^{2}$\\\hline
$\mathfrak{n}_{7.62}$\newline {\scriptsize $1 \oplus 1,3,4,6_B$} & \mbox{$[e_{3},e_{7}]=e_{2}$}, \mbox{$[e_{4},e_{6}]=e_{2}$}, \mbox{$[e_{5},e_{6}]=e_{4}$}, \mbox{$[e_{5},e_{7}]=e_{2}$}, \mbox{$[e_{6},e_{7}]=e_{5}$} & $e_{1}$, $e_{2}$, $2 e_{2} e_{3}- 2 e_{2} e_{5}+e_{4}^{2}$\\\hline
$\mathfrak{n}_{7.63}$\newline {\scriptsize $1 \oplus 1,3,4,6_C$} & \mbox{$[e_{3},e_{6}]=e_{2}$}, \mbox{$[e_{4},e_{7}]=e_{2}$}, \mbox{$[e_{5},e_{6}]=e_{4}$}, \mbox{$[e_{5},e_{7}]=e_{3}$}, \mbox{$[e_{6},e_{7}]=e_{5}$} & $e_{1}$, $e_{2}$, $e_{2} e_{5}- e_{3} e_{4}$\\\hline
$\mathfrak{n}_{7.64}$\newline {\scriptsize $2,4,5,7_A$} & \mbox{$[e_{3},e_{6}]=e_{1}$}, \mbox{$[e_{4},e_{6}]=e_{2}$}, \mbox{$[e_{5},e_{6}]=e_{4}$}, \mbox{$[e_{6},e_{7}]=e_{5}$} & $e_{1}$, $e_{2}$, $e_{1} e_{4}- e_{2} e_{3}$, $- 2 e_{1}^{2} e_{5}+2 e_{1} e_{3} e_{4}- e_{2} e_{3}^{2}$, $- 3 e_{1} e_{2} e_{7}- e_{1} e_{4} e_{5}- 2 e_{2} e_{3} e_{5} +e_{3} e_{4}^{2}$\\\hline
$\mathfrak{n}_{7.65}$\newline {\scriptsize $2,4,5,7_B$} & \mbox{$[e_{3},e_{7}]=e_{2}$}, \mbox{$[e_{4},e_{6}]=e_{1}$}, \mbox{$[e_{5},e_{6}]=e_{4}$}, \mbox{$[e_{6},e_{7}]=e_{5}$} & $e_{1}$, $e_{2}$, $2 e_{1} e_{5}- e_{4}^{2}$\\\hline
$\mathfrak{n}_{7.66}$\newline {\scriptsize $2,4,5,7_C$} & \mbox{$[e_{3},e_{6}]=e_{1}$}, \mbox{$[e_{3},e_{7}]=e_{2}$}, \mbox{$[e_{4},e_{6}]=e_{2}$}, \mbox{$[e_{5},e_{6}]=e_{4}$}, \mbox{$[e_{6},e_{7}]=e_{5}$} & $e_{1}$, $e_{2}$, $2 e_{2} e_{5}- e_{4}^{2}$\\\hline
$\mathfrak{n}_{7.67}$\newline {\scriptsize $2,4,5,7_D$} & \mbox{$[e_{3},e_{6}]=e_{1}$}, \mbox{$[e_{3},e_{7}]=e_{2}$}, \mbox{$[e_{4},e_{6}]=e_{2}$}, \mbox{$[e_{5},e_{6}]=e_{4}$}, \mbox{$[e_{5},e_{7}]=e_{2}$}, \mbox{$[e_{6},e_{7}]=e_{5}$} & $e_{1}$, $e_{2}$, $- 2 e_{1} e_{4}+2 e_{2} e_{3}- 2 e_{2} e_{5}+e_{4}^{2}$\\\hline
$\mathfrak{n}_{7.68}$\newline {\scriptsize $2,4,5,7_E$} & \mbox{$[e_{3},e_{7}]=e_{2}$}, \mbox{$[e_{4},e_{6}]=e_{1}$}, \mbox{$[e_{5},e_{6}]=e_{4}$}, \mbox{$[e_{5},e_{7}]=e_{2}$}, \mbox{$[e_{6},e_{7}]=e_{5}$} & $e_{1}$, $e_{2}$, $2 e_{1} e_{3}- 2 e_{1} e_{5}+e_{4}^{2}$\\\hline
$\mathfrak{n}_{7.69}$\newline {\scriptsize $2,4,5,7_F$} & \mbox{$[e_{3},e_{6}]=e_{1}$}, \mbox{$[e_{4},e_{6}]=e_{2}$}, \mbox{$[e_{5},e_{6}]=e_{4}$}, \mbox{$[e_{5},e_{7}]=e_{2}$}, \mbox{$[e_{6},e_{7}]=e_{5}$} & $e_{1}$, $e_{2}$, $e_{1} e_{4}- e_{2} e_{3}$\\\hline
$\mathfrak{n}_{7.70}$\newline {\scriptsize $2,4,5,7_G$} & \mbox{$[e_{3},e_{6}]=e_{2}$}, \mbox{$[e_{4},e_{6}]=e_{1}$}, \mbox{$[e_{5},e_{6}]=e_{4}$}, \mbox{$[e_{5},e_{7}]=e_{2}$}, \mbox{$[e_{6},e_{7}]=e_{5}$} & $e_{1}$, $e_{2}$, $e_{1} e_{3}- e_{2} e_{4}$\\\hline
$\mathfrak{n}_{7.71}$\newline {\scriptsize $2,4,5,7_H$} & \mbox{$[e_{3},e_{7}]=e_{1}$}, \mbox{$[e_{4},e_{6}]=e_{1}$}, \mbox{$[e_{5},e_{6}]=e_{4}$}, \mbox{$[e_{5},e_{7}]=e_{2}$}, \mbox{$[e_{6},e_{7}]=e_{5}$} & $e_{1}$, $e_{2}$, $2 e_{1} e_{5}- 2 e_{2} e_{3}- e_{4}^{2}$\\\hline
$\mathfrak{n}_{7.72}$\newline {\scriptsize $2,4,5,7_I$} & \mbox{$[e_{3},e_{7}]=e_{1}$}, \mbox{$[e_{4},e_{6}]=e_{2}$}, \mbox{$[e_{5},e_{6}]=e_{4}$}, \mbox{$[e_{5},e_{7}]=e_{2}$}, \mbox{$[e_{6},e_{7}]=e_{5}$} & $e_{1}$, $e_{2}$, $2 e_{1} e_{2} e_{5}- e_{1} e_{4}^{2}- 2 e_{2}^{2} e_{3}$\\\hline
$\mathfrak{n}_{7.73}$\newline {\scriptsize $2,4,5,7_J$} & \mbox{$[e_{3},e_{7}]=e_{1}$}, \mbox{$[e_{4},e_{6}]=e_{1} + e_{2}$}, \mbox{$[e_{5},e_{6}]=e_{4}$}, \mbox{$[e_{5},e_{7}]=e_{2}$}, \mbox{$[e_{6},e_{7}]=e_{5}$} & $e_{1}$, $e_{2}$, $- 2 e_{1}^{2} e_{5}+2 e_{1} e_{2} e_{3}- 2 e_{1} e_{2} e_{5} +e_{1} e_{4}^{2}+2 e_{2}^{2} e_{3}$\\\hline
$\mathfrak{n}_{7.74}$\newline {\scriptsize $2,4,5,7_K$} & \mbox{$[e_{3},e_{6}]=e_{2}$}, \mbox{$[e_{3},e_{7}]=e_{1}$}, \mbox{$[e_{4},e_{6}]=e_{1}$}, \mbox{$[e_{5},e_{6}]=e_{4}$}, \mbox{$[e_{5},e_{7}]=e_{2}$}, \mbox{$[e_{6},e_{7}]=e_{5}$} & $e_{1}$, $e_{2}$, $- 2 e_{1}^{2} e_{5}+2 e_{1} e_{2} e_{3}+e_{1} e_{4}^{2} - 2 e_{2}^{2} e_{4}$\\\hline
$\mathfrak{n}_{7.75}$\newline {\scriptsize $2,4,5,7_L$} & \mbox{$[e_{3},e_{7}]=e_{1}$}, \mbox{$[e_{4},e_{6}]=e_{2}$}, \mbox{$[e_{5},e_{6}]=e_{4}$}, \mbox{$[e_{5},e_{7}]=e_{3}$}, \mbox{$[e_{6},e_{7}]=e_{5}$} & $e_{1}$, $e_{2}$, $2 e_{1} e_{2} e_{5}- e_{1} e_{4}^{2}- e_{2} e_{3}^{2}$\\\hline
$\mathfrak{n}_{7.76}$\newline {\scriptsize $2,4,5,7_M$} & \mbox{$[e_{3},e_{6}]=e_{2}$}, \mbox{$[e_{4},e_{6}]=e_{1}$}, \mbox{$[e_{4},e_{7}]=e_{2}$}, \mbox{$[e_{5},e_{6}]=e_{4}$}, \mbox{$[e_{5},e_{7}]=e_{3}$}, \mbox{$[e_{6},e_{7}]=e_{5}$} & $e_{1}$, $e_{2}$, $- e_{1} e_{3}^{2}- 2 e_{2}^{2} e_{5}+2 e_{2} e_{3} e_{4}$\\\hline
$\mathfrak{n}_{7.77}$\newline {\scriptsize $1 \oplus 1,2,4,6$} & \mbox{$[e_{3},e_{6}]=e_{2}$}, \mbox{$[e_{4},e_{5}]=- e_{2}$}, \mbox{$[e_{4},e_{7}]=e_{3}$}, \mbox{$[e_{5},e_{6}]=e_{3}$}, \mbox{$[e_{6},e_{7}]=e_{5}$} & $e_{1}$, $e_{2}$, $6 e_{2}^{2} e_{7}+6 e_{2} e_{3} e_{5}- 2 e_{3}^{3}$\\\hline
$\mathfrak{n}_{7.78}$\newline {\scriptsize $2,3,5,7_A$} & \mbox{$[e_{3},e_{6}]=e_{2}$}, \mbox{$[e_{4},e_{5}]=- e_{2}$}, \mbox{$[e_{4},e_{7}]=e_{1} + e_{3}$}, \mbox{$[e_{5},e_{6}]=e_{3}$}, \mbox{$[e_{6},e_{7}]=e_{5}$} & $e_{1}$, $e_{2}$, $6 e_{1} e_{2} e_{5}- 3 e_{1} e_{3}^{2}+6 e_{2}^{2} e_{7} +6 e_{2} e_{3} e_{5}- 2 e_{3}^{3}$\\\hline
$\mathfrak{n}_{7.79}$\newline {\scriptsize $2,3,5,7_B$} & \mbox{$[e_{3},e_{6}]=e_{2}$}, \mbox{$[e_{4},e_{5}]=- e_{2}$}, \mbox{$[e_{4},e_{6}]=e_{1}$}, \mbox{$[e_{4},e_{7}]=e_{3}$}, \mbox{$[e_{5},e_{6}]=e_{3}$}, \mbox{$[e_{6},e_{7}]=e_{5}$} & $e_{1}$, $e_{2}$, $6 e_{2}^{2} e_{7}+6 e_{2} e_{3} e_{5}- 2 e_{3}^{3}$\\\hline
$\mathfrak{n}_{7.80}$\newline {\scriptsize $2,3,5,7_C$} & \mbox{$[e_{3},e_{6}]=e_{2}$}, \mbox{$[e_{4},e_{5}]=- e_{2}$}, \mbox{$[e_{4},e_{7}]=e_{3}$}, \mbox{$[e_{5},e_{6}]=e_{3}$}, \mbox{$[e_{5},e_{7}]=e_{1}$}, \mbox{$[e_{6},e_{7}]=e_{5}$} & $e_{1}$, $e_{2}$, $- 3 e_{1} e_{2} e_{4}+3 e_{2}^{2} e_{7}+3 e_{2} e_{3} e_{5} - e_{3}^{3}$\\\hline
$\mathfrak{n}_{7.81}$\newline {\scriptsize $2,3,5,7_D$} & \mbox{$[e_{3},e_{6}]=e_{2}$}, \mbox{$[e_{4},e_{5}]=- e_{2}$}, \mbox{$[e_{4},e_{7}]=e_{1} + e_{3}$}, \mbox{$[e_{5},e_{6}]=e_{3}$}, \mbox{$[e_{5},e_{7}]=e_{1}$}, \mbox{$[e_{6},e_{7}]=e_{5}$} & $e_{1}$, $e_{2}$, $6 e_{1} e_{2} e_{4}- 6 e_{1} e_{2} e_{5}+3 e_{1} e_{3}^{2} - 6 e_{2}^{2} e_{7}- 6 e_{2} e_{3} e_{5}+2 e_{3}^{3}$\\\hline
$\mathfrak{n}_{7.82}$\newline {\scriptsize $1 \oplus 1,2,3,4,6_A$} & \mbox{$[e_{3},e_{6}]=e_{2}$}, \mbox{$[e_{4},e_{6}]=e_{3}$}, \mbox{$[e_{5},e_{6}]=e_{4}$}, \mbox{$[e_{6},e_{7}]=e_{5}$} & $e_{1}$, $e_{2}$, $2 e_{2} e_{4}- e_{3}^{2}$, $- 8 e_{2}^{3} e_{7}- 8 e_{2}^{2} e_{3} e_{5} +4 e_{2} e_{3}^{2} e_{4}- e_{3}^{4}$, $- 3 e_{2}^{2} e_{5}+3 e_{2} e_{3} e_{4}- e_{3}^{3}$\\\hline
$\mathfrak{n}_{7.83}$\newline {\scriptsize $1 \oplus 1,2,3,4,6_B$} & \mbox{$[e_{3},e_{6}]=e_{2}$}, \mbox{$[e_{4},e_{6}]=e_{3}$}, \mbox{$[e_{5},e_{6}]=e_{4}$}, \mbox{$[e_{5},e_{7}]=e_{2}$}, \mbox{$[e_{6},e_{7}]=e_{5}$} & $e_{1}$, $e_{2}$, $2 e_{2} e_{4}- e_{3}^{2}$\\\hline
$\mathfrak{n}_{7.84}$\newline {\scriptsize $1 \oplus 1,2,3,4,6_C$} & \mbox{$[e_{3},e_{7}]=e_{2}$}, \mbox{$[e_{4},e_{5}]=e_{2}$}, \mbox{$[e_{4},e_{6}]=e_{3}$}, \mbox{$[e_{5},e_{6}]=e_{4}$}, \mbox{$[e_{6},e_{7}]=e_{5}$} & $e_{1}$, $e_{2}$, $2 e_{2} e_{6}- 2 e_{3} e_{5}+e_{4}^{2}$\\\hline
$\mathfrak{n}_{7.85}$\newline {\scriptsize $1 \oplus 1,2,3,4,6_D$} & \mbox{$[e_{3},e_{6}]=e_{2}$}, \mbox{$[e_{4},e_{6}]=e_{3}$}, \mbox{$[e_{4},e_{7}]=e_{2}$}, \mbox{$[e_{5},e_{6}]=e_{4}$}, \mbox{$[e_{5},e_{7}]=e_{3}$}, \mbox{$[e_{6},e_{7}]=e_{5}$} & $e_{1}$, $e_{2}$, $- 3 e_{2}^{2} e_{5}+3 e_{2} e_{3} e_{4}- e_{3}^{3}$\\\hline
$\mathfrak{n}_{7.86}$\newline {\scriptsize $1 \oplus 1,2,3,4,6_E$} & \mbox{$[e_{3},e_{7}]=e_{2}$}, \mbox{$[e_{4},e_{5}]=e_{2}$}, \mbox{$[e_{4},e_{6}]=e_{3}$}, \mbox{$[e_{5},e_{6}]=e_{4}$}, \mbox{$[e_{5},e_{7}]=e_{3}$}, \mbox{$[e_{6},e_{7}]=e_{5}$} & $e_{1}$, $e_{2}$, $6 e_{2}^{2} e_{6}- 6 e_{2} e_{3} e_{5}+3 e_{2} e_{4}^{2} +2 e_{3}^{3}$\\\hline
$\mathfrak{n}_{7.87}$\newline {\scriptsize $2,3,4,5,7_A$} & \mbox{$[e_{3},e_{6}]=e_{2}$}, \mbox{$[e_{4},e_{6}]=e_{3}$}, \mbox{$[e_{5},e_{6}]=e_{4}$}, \mbox{$[e_{5},e_{7}]=e_{1}$}, \mbox{$[e_{6},e_{7}]=e_{5}$} & $e_{1}$, $e_{2}$, $2 e_{2} e_{4}- e_{3}^{2}$\\\hline
$\mathfrak{n}_{7.88}$\newline {\scriptsize $2,3,4,5,7_B$} & \mbox{$[e_{3},e_{7}]=e_{2}$}, \mbox{$[e_{4},e_{5}]=e_{2}$}, \mbox{$[e_{4},e_{6}]=e_{3}$}, \mbox{$[e_{5},e_{6}]=e_{4}$}, \mbox{$[e_{5},e_{7}]=e_{1}$}, \mbox{$[e_{6},e_{7}]=e_{5}$} & $e_{1}$, $e_{2}$, $e_{1} e_{3}^{2}+2 e_{2}^{2} e_{6}- 2 e_{2} e_{3} e_{5} +e_{2} e_{4}^{2}$\\\hline
$\mathfrak{n}_{7.89}$\newline {\scriptsize $2,3,4,5,7_C$} & \mbox{$[e_{3},e_{6}]=e_{2}$}, \mbox{$[e_{3},e_{7}]=e_{1}$}, \mbox{$[e_{4},e_{5}]=e_{1}$}, \mbox{$[e_{4},e_{6}]=e_{3}$}, \mbox{$[e_{5},e_{6}]=e_{4}$}, \mbox{$[e_{6},e_{7}]=e_{5}$} & $e_{1}$, $e_{2}$, $2 e_{1} e_{6}- 2 e_{2} e_{7}- 2 e_{3} e_{5}+e_{4}^{2}$\\\hline
$\mathfrak{n}_{7.90}$\newline {\scriptsize $2,3,4,5,7_D$} & \mbox{$[e_{3},e_{6}]=e_{2}$}, \mbox{$[e_{3},e_{7}]=e_{1}$}, \mbox{$[e_{4},e_{5}]=e_{1}$}, \mbox{$[e_{4},e_{6}]=e_{3}$}, \mbox{$[e_{5},e_{6}]=e_{4}$}, \mbox{$[e_{5},e_{7}]=e_{2}$}, \mbox{$[e_{6},e_{7}]=e_{5}$} & $e_{1}$, $e_{2}$, $2 e_{1}^{2} e_{6}- 2 e_{1} e_{2} e_{7}- 2 e_{1} e_{3} e_{5} +e_{1} e_{4}^{2}- 2 e_{2}^{2} e_{4}+e_{2} e_{3}^{2}$\\\hline
$\mathfrak{n}_{7.91}$\newline {\scriptsize $2,3,4,5,7_E$} & \mbox{$[e_{3},e_{6}]=e_{2}$}, \mbox{$[e_{4},e_{6}]=e_{3}$}, \mbox{$[e_{4},e_{7}]=e_{2}$}, \mbox{$[e_{5},e_{6}]=e_{4}$}, \mbox{$[e_{5},e_{7}]=e_{1} + e_{3}$}, \mbox{$[e_{6},e_{7}]=e_{5}$} & $e_{1}$, $e_{2}$, $- 6 e_{1} e_{2} e_{4}+3 e_{1} e_{3}^{2}+6 e_{2}^{2} e_{5} - 6 e_{2} e_{3} e_{4}+2 e_{3}^{3}$\\\hline
$\mathfrak{n}_{7.92}$\newline {\scriptsize $2,3,4,5,7_F$} & \mbox{$[e_{3},e_{7}]=e_{2}$}, \mbox{$[e_{4},e_{5}]=e_{2}$}, \mbox{$[e_{4},e_{6}]=e_{3}$}, \mbox{$[e_{5},e_{6}]=e_{4}$}, \mbox{$[e_{5},e_{7}]=e_{1} + e_{3}$}, \mbox{$[e_{6},e_{7}]=e_{5}$} & $e_{1}$, $e_{2}$, $3 e_{1} e_{3}^{2}+6 e_{2}^{2} e_{6}- 6 e_{2} e_{3} e_{5} +3 e_{2} e_{4}^{2}+2 e_{3}^{3}$\\\hline
$\mathfrak{n}_{7.93}$\newline {\scriptsize $2,3,4,5,7_G$} & \mbox{$[e_{3},e_{6}]=e_{2}$}, \mbox{$[e_{3},e_{7}]=e_{1}$}, \mbox{$[e_{4},e_{5}]=e_{1}$}, \mbox{$[e_{4},e_{6}]=e_{3}$}, \mbox{$[e_{4},e_{7}]=e_{2}$}, \mbox{$[e_{5},e_{6}]=e_{4}$}, \mbox{$[e_{5},e_{7}]=e_{3}$}, \mbox{$[e_{6},e_{7}]=e_{5}$} & $e_{1}$, $e_{2}$, $6 e_{1}^{2} e_{6}- 6 e_{1} e_{2} e_{7}- 6 e_{1} e_{3} e_{5} +3 e_{1} e_{4}^{2}+6 e_{2}^{2} e_{5}- 6 e_{2} e_{3} e_{4} +2 e_{3}^{3}$\\\hline

$\mathfrak{n}_{7.94}$\newline {\scriptsize $1,7$} & \mbox{$[e_{2},e_{3}]=e_{1}$}, \mbox{$[e_{4},e_{5}]=e_{1}$}, \mbox{$[e_{6},e_{7}]=e_{1}$} & $e_{1}$\\\hline
$\mathfrak{n}_{7.95}$\newline {\scriptsize $1,5,7$} & \mbox{$[e_{2},e_{3}]=e_{1}$}, \mbox{$[e_{4},e_{6}]=e_{1}$}, \mbox{$[e_{5},e_{7}]=e_{1}$}, \mbox{$[e_{6},e_{7}]=e_{5}$} & $e_{1}$\\\hline
$\mathfrak{n}_{7.96}$\newline {\scriptsize $1,4,7_{A}$} & \mbox{$[e_{2},e_{7}]=e_{1}$}, \mbox{$[e_{3},e_{6}]=e_{1}$}, \mbox{$[e_{4},e_{5}]=e_{1}$}, \mbox{$[e_{5},e_{7}]=e_{3}$}, \mbox{$[e_{6},e_{7}]=e_{4}$} & $e_{1}$\\\hline
$\mathfrak{n}_{7.97}$\newline {\scriptsize $1,4,7_{B}$} & \mbox{$[e_{2},e_{6}]=e_{1}$}, \mbox{$[e_{3},e_{5}]=e_{1}$}, \mbox{$[e_{4},e_{7}]=e_{1}$}, \mbox{$[e_{5},e_{7}]=e_{3}$}, \mbox{$[e_{6},e_{7}]=e_{4}$} & $e_{1}$\\\hline
$\mathfrak{n}_{7.98}$\newline {\scriptsize $1,4,7_{C}$} & \mbox{$[e_{2},e_{6}]=\tfrac12 e_{1}$}, \mbox{$[e_{3},e_{7}]=- e_{1}$}, \mbox{$[e_{4},e_{5}]=\tfrac12 e_{1}$}, \mbox{$[e_{5},e_{6}]=e_{3}$}, \mbox{$[e_{5},e_{7}]=- e_{2}$}, \mbox{$[e_{6},e_{7}]=e_{4}$} & $e_{1}$\\\hline
$\mathfrak{n}_{7.99}$\newline {\scriptsize $1,4,7_{D}$} & \mbox{$[e_{2},e_{5}]=e_{1}$}, \mbox{$[e_{2},e_{6}]=\tfrac12 e_{1}$}, \mbox{$[e_{3},e_{7}]=- e_{1}$}, \mbox{$[e_{4},e_{5}]=\tfrac12 e_{1}$}, \mbox{$[e_{5},e_{6}]=e_{3}$}, \mbox{$[e_{5},e_{7}]=- e_{2}$}, \mbox{$[e_{6},e_{7}]=e_{4}$} & $e_{1}$\\\hline

$\mathfrak{n}_{7.100}(\xi)$\newline {\scriptsize $1,4,7_{E}$}\newline {\scriptsize $\xi\notin\{0,1,2,-1,\tfrac12\}$} & \mbox{$[e_{2},e_{6}]=\xi e_{1}$}, \mbox{$[e_{3},e_{7}]=- e_{1}$}, \mbox{$[e_{4},e_{5}]=(1-\xi)e_{1}$}, \mbox{$[e_{5},e_{6}]=e_{3}$}, \mbox{$[e_{5},e_{7}]=- e_{2}$}, \mbox{$[e_{6},e_{7}]=e_{4}$} & $e_{1}$\\\hline
$\mathfrak{n}_{7.101}$\newline {\scriptsize $1,4,5,7_{A}$} & \mbox{$[e_{2},e_{3}]=e_{1}$}, \mbox{$[e_{4},e_{7}]=e_{1}$}, \mbox{$[e_{5},e_{7}]=e_{4}$}, \mbox{$[e_{6},e_{7}]=e_{5}$} & $e_{1}$, $- 2 e_{1} e_{5} + e_{4}^{2}$, $3 e_{1}^{2} e_{6} - 3 e_{1} e_{4} e_{5} + e_{4}^{3}$\\\hline
$\mathfrak{n}_{7.102}$\newline {\scriptsize $1,4,5,7_{B}$} & \mbox{$[e_{2},e_{3}]=e_{1}$}, \mbox{$[e_{4},e_{7}]=e_{1}$}, \mbox{$[e_{5},e_{6}]=e_{1}$}, \mbox{$[e_{5},e_{7}]=e_{4}$}, \mbox{$[e_{6},e_{7}]=e_{5}$} & $e_{1}$\\\hline
$\mathfrak{n}_{7.103}$\newline {\scriptsize $1,3,7_{A}$} & \mbox{$[e_{2},e_{5}]=e_{1}$}, \mbox{$[e_{3},e_{7}]=e_{1}$}, \mbox{$[e_{4},e_{5}]=e_{2}$}, \mbox{$[e_{6},e_{7}]=e_{3}$} & $e_{1}$, $- 2 e_{1} e_{4} + e_{2}^{2}$, $- 2 e_{1} e_{6} + e_{3}^{2}$\\\hline
$\mathfrak{n}_{7.104}$\newline {\scriptsize $1,3,7_{B}$} & \mbox{$[e_{2},e_{5}]=e_{1}$}, \mbox{$[e_{3},e_{7}]=e_{1}$}, \mbox{$[e_{4},e_{5}]=e_{2}$}, \mbox{$[e_{4},e_{6}]=e_{1}$}, \mbox{$[e_{6},e_{7}]=e_{3}$} & $e_{1}$\\\hline
$\mathfrak{n}_{7.105}$\newline {\scriptsize $1,3,7_{C}$} & \mbox{$[e_{2},e_{4}]=- e_{1}$}, \mbox{$[e_{3},e_{7}]=e_{1}$}, \mbox{$[e_{4},e_{5}]=e_{3}$}, \mbox{$[e_{5},e_{7}]=e_{2}$}, \mbox{$[e_{6},e_{7}]=e_{3}$} & $e_{1}$, $- e_{1} e_{5} + e_{2} e_{3}$, $- 2 e_{1} e_{6} + e_{3}^{2}$\\\hline
$\mathfrak{n}_{7.106}$\newline {\scriptsize $1,3,7_{D}$} & \mbox{$[e_{2},e_{4}]=- e_{1}$}, \mbox{$[e_{3},e_{7}]=e_{1}$}, \mbox{$[e_{4},e_{5}]=e_{3}$}, \mbox{$[e_{5},e_{6}]=e_{1}$}, \mbox{$[e_{5},e_{7}]=e_{2}$}, \mbox{$[e_{6},e_{7}]=e_{3}$} & $e_{1}$\\\hline
$\mathfrak{n}_{7.107}$\newline {\scriptsize $1,3,5,7_{A}$} & \mbox{$[e_{2},e_{6}]=e_{1}$}, \mbox{$[e_{3},e_{7}]=e_{1}$}, \mbox{$[e_{4},e_{5}]=e_{1}$}, \mbox{$[e_{4},e_{6}]=e_{3}$}, \mbox{$[e_{5},e_{7}]=e_{3}$}, \mbox{$[e_{6},e_{7}]=e_{5}$} & $e_{1}$\\\hline
$\mathfrak{n}_{7.108}$\newline {\scriptsize $1,3,5,7_{B}$} & \mbox{$[e_{2},e_{4}]=e_{1}$}, \mbox{$[e_{3},e_{7}]=e_{1}$}, \mbox{$[e_{4},e_{5}]=e_{1}$}, \mbox{$[e_{4},e_{6}]=e_{3}$}, \mbox{$[e_{5},e_{7}]=e_{3}$}, \mbox{$[e_{6},e_{7}]=e_{5}$} & $e_{1}$, $- 2 e_{1} e_{2} - 2 e_{1} e_{5} + e_{3}^{2}$, $3 e_{1}^{2} e_{6} - 3 e_{1} e_{3} e_{5} + e_{3}^{3}$\\\hline
$\mathfrak{n}_{7.109}$\newline {\scriptsize $1,3,5,7_{C}$} & \mbox{$[e_{2},e_{4}]=e_{1}$}, \mbox{$[e_{3},e_{7}]=e_{1}$}, \mbox{$[e_{4},e_{5}]=e_{1}$}, \mbox{$[e_{4},e_{6}]=e_{3}$}, \mbox{$[e_{5},e_{6}]=e_{1}$}, \mbox{$[e_{5},e_{7}]=e_{3}$}, \mbox{$[e_{6},e_{7}]=e_{5}$} & $e_{1}$\\\hline
$\mathfrak{n}_{7.110}$\newline {\scriptsize $1,3,5,7_{D}$} & \mbox{$[e_{2},e_{6}]=e_{1}$}, \mbox{$[e_{3},e_{7}]=e_{1}$}, \mbox{$[e_{4},e_{5}]=- e_{1}$}, \mbox{$[e_{4},e_{7}]=e_{2}$}, \mbox{$[e_{5},e_{7}]=e_{3}$}, \mbox{$[e_{6},e_{7}]=e_{5}$} & $e_{1}$\\\hline
$\mathfrak{n}_{7.111}$\newline {\scriptsize $1,3,5,7_{E}$} & \mbox{$[e_{2},e_{4}]=e_{1}$}, \mbox{$[e_{3},e_{7}]=e_{1}$}, \mbox{$[e_{4},e_{7}]=e_{2}$}, \mbox{$[e_{5},e_{7}]=e_{3}$}, \mbox{$[e_{6},e_{7}]=e_{5}$} & $e_{1}$, $- 2 e_{1} e_{5} + e_{3}^{2}$, $3 e_{1}^{2} e_{6} - 3 e_{1} e_{3} e_{5} + e_{3}^{3}$\\\hline
$\mathfrak{n}_{7.112}$\newline {\scriptsize $1,3,5,7_{F}$} & \mbox{$[e_{2},e_{4}]=e_{1}$}, \mbox{$[e_{3},e_{7}]=e_{1}$}, \mbox{$[e_{4},e_{7}]=e_{2}$}, \mbox{$[e_{5},e_{6}]=e_{1}$}, \mbox{$[e_{5},e_{7}]=e_{3}$}, \mbox{$[e_{6},e_{7}]=e_{5}$} & $e_{1}$\\\hline
$\mathfrak{n}_{7.113}$\newline {\scriptsize $1,3,5,7_{G}$} & \mbox{$[e_{2},e_{6}]=e_{1}$}, \mbox{$[e_{3},e_{7}]=e_{1}$}, \mbox{$[e_{4},e_{6}]=e_{2}$}, \mbox{$[e_{5},e_{7}]=e_{3}$}, \mbox{$[e_{6},e_{7}]=e_{5}$} & $e_{1}$, $- 2 e_{1} e_{4} + e_{2}^{2}$, $- 2 e_{1} e_{5} + e_{3}^{2}$\\\hline
$\mathfrak{n}_{7.114}$\newline {\scriptsize $1,3,5,7_{H}$} & \mbox{$[e_{2},e_{6}]=e_{1}$}, \mbox{$[e_{2},e_{7}]=e_{1}$}, \mbox{$[e_{3},e_{7}]=e_{1}$}, \mbox{$[e_{4},e_{5}]=e_{1}$}, \mbox{$[e_{4},e_{6}]=e_{2}$}, \mbox{$[e_{5},e_{7}]=e_{3}$}, \mbox{$[e_{6},e_{7}]=e_{5}$} & $e_{1}$\\\hline
$\mathfrak{n}_{7.115}$\newline {\scriptsize $1,3,5,7_{I}$} & \mbox{$[e_{2},e_{4}]=e_{1}$}, \mbox{$[e_{3},e_{7}]=e_{1}$}, \mbox{$[e_{4},e_{6}]=e_{2}$}, \mbox{$[e_{5},e_{7}]=e_{3}$}, \mbox{$[e_{6},e_{7}]=e_{5}$} & $e_{1}$, $- 2 e_{1} e_{5} + e_{3}^{2}$, $6 e_{1}^{2} e_{6} + 3 e_{1} e_{2}^{2} - 6 e_{1} e_{3} e_{5} + 2 e_{3}^{3}$\\\hline
$\mathfrak{n}_{7.116}$\newline {\scriptsize $1,3,5,7_{J}$} & \mbox{$[e_{2},e_{4}]=e_{1}$}, \mbox{$[e_{3},e_{7}]=e_{1}$}, \mbox{$[e_{4},e_{6}]=e_{2}$}, \mbox{$[e_{5},e_{6}]=e_{1}$}, \mbox{$[e_{5},e_{7}]=e_{3}$}, \mbox{$[e_{6},e_{7}]=e_{5}$} & $e_{1}$\\\hline
$\mathfrak{n}_{7.117}$\newline {\scriptsize $1,3,5,7_{K}$} & \mbox{$[e_{2},e_{6}]=\tfrac12 e_{1}$}, \mbox{$[e_{3},e_{7}]=e_{1}$}, \mbox{$[e_{4},e_{5}]=\tfrac12 e_{1}$}, \mbox{$[e_{4},e_{6}]=e_{3}$}, \mbox{$[e_{4},e_{7}]=e_{2}$}, \mbox{$[e_{5},e_{7}]=e_{3}$}, \mbox{$[e_{6},e_{7}]=e_{5}$} & $e_{1}$\\\hline
$\mathfrak{n}_{7.118}$\newline {\scriptsize $1,3,5,7_{L}$} & \mbox{$[e_{2},e_{6}]=\tfrac12 e_{1}$}, \mbox{$[e_{3},e_{7}]=e_{1}$}, \mbox{$[e_{4},e_{5}]=\tfrac12 e_{1}$}, \mbox{$[e_{4},e_{6}]=e_{3}$}, \mbox{$[e_{4},e_{7}]=e_{2}$}, \mbox{$[e_{5},e_{6}]=e_{1}$}, \mbox{$[e_{5},e_{7}]=e_{3}$}, \mbox{$[e_{6},e_{7}]=e_{5}$} & $e_{1}$\\\hline

$\mathfrak{n}_{7.119}(\xi)$\newline {\scriptsize $1,3,5,7_{M}$}\newline {\scriptsize $\xi\notin\{0,\frac12\}$} & \mbox{$[e_{2},e_{6}]=\xi e_{1}$}, \mbox{$[e_{3},e_{7}]=e_{1}$}, \mbox{$[e_{4},e_{5}]=(1-\xi)e_{1}$}, \mbox{$[e_{4},e_{6}]=e_{3}$}, \mbox{$[e_{4},e_{7}]=e_{2}$}, \mbox{$[e_{5},e_{7}]=e_{3}$}, \mbox{$[e_{6},e_{7}]=e_{5}$} & $e_{1}$ ($\xi\ne 1$);\newline $e_{1}$, $e_2e_3-e_1e_4$, $e_3^2-2e_1e_5$ ($\xi=1$)\\\hline

$\mathfrak{n}_{7.120}(\xi)$\newline {\scriptsize $1,3,5,7_{N}$}\newline {\scriptsize $\xi\in\mathbb C$} & \mbox{$[e_{2},e_{4}]=e_{1}$}, \mbox{$[e_{3},e_{7}]=e_{1}$}, \mbox{$[e_{4},e_{5}]=e_{1}$}, \mbox{$[e_{4},e_{6}]=e_{3}$}, \mbox{$[e_{4},e_{7}]=e_{2}$}, \mbox{$[e_{5},e_{6}]=\xi e_{1}$}, \mbox{$[e_{5},e_{7}]=e_{3}$}, \mbox{$[e_{6},e_{7}]=e_{5}$} & $e_{1}$ ($\xi\ne 0$);\newline $e_{1}$, $- 2 e_{1} e_{2} - 2 e_{1} e_{5} + e_{3}^{2}$, $3 e_{1}^{2} e_{6} - 3 e_{1} e_{3} e_{5} + e_{3}^{3}$ ($\xi=0$)\\\hline
$\mathfrak{n}_{7.121}$\newline {\scriptsize $1,3,5,7_{O}$} & \mbox{$[e_{2},e_{7}]=e_{1}$}, \mbox{$[e_{3},e_{6}]=e_{1}$}, \mbox{$[e_{4},e_{7}]=e_{2}$}, \mbox{$[e_{5},e_{6}]=e_{2}$}, \mbox{$[e_{5},e_{7}]=e_{3}$}, \mbox{$[e_{6},e_{7}]=e_{5}$} & $e_{1}$, $- 2 e_{1} e_{4} + e_{2}^{2}$, $- e_{1} e_{5} + e_{2} e_{3}$\\\hline
$\mathfrak{n}_{7.122}$\newline {\scriptsize $1,3,5,7_{P}$} & \mbox{$[e_{2},e_{6}]=e_{1}$}, \mbox{$[e_{3},e_{7}]=e_{1}$}, \mbox{$[e_{4},e_{5}]=- e_{1}$}, \mbox{$[e_{4},e_{7}]=e_{2}$}, \mbox{$[e_{5},e_{6}]=e_{2}$}, \mbox{$[e_{5},e_{7}]=e_{3}$}, \mbox{$[e_{6},e_{7}]=e_{5}$} & $e_{1}$\\\hline
$\mathfrak{n}_{7.123}$\newline {\scriptsize $1,3,5,7_{Q}$} & \mbox{$[e_{2},e_{6}]=e_{1}$}, \mbox{$[e_{3},e_{7}]=e_{1}$}, \mbox{$[e_{4},e_{7}]=e_{3}$}, \mbox{$[e_{5},e_{6}]=e_{2}$}, \mbox{$[e_{6},e_{7}]=e_{4} + e_{5}$} & $e_{1}$, $- 2 e_{1} e_{5} + e_{2}^{2}$, $- 2 e_{1} e_{4} + e_{3}^{2}$\\\hline
$\mathfrak{n}_{7.124}$\newline {\scriptsize $1,3,5,7_{R}$} & \mbox{$[e_{2},e_{7}]=e_{1}$}, \mbox{$[e_{3},e_{6}]=e_{1}$}, \mbox{$[e_{4},e_{5}]=- e_{1}$}, \mbox{$[e_{4},e_{7}]=e_{3}$}, \mbox{$[e_{5},e_{6}]=e_{2}$}, \mbox{$[e_{6},e_{7}]=e_{4} + e_{5}$} & $e_{1}$\\\hline

$\mathfrak{n}_{7.125}(\xi)$\newline {\scriptsize $1,3,5,7_{S}$}\newline {\scriptsize $\xi\ne1$} & \mbox{$[e_{2},e_{6}]=\xi e_{1}$}, \mbox{$[e_{2},e_{7}]=e_{1}$}, \mbox{$[e_{3},e_{6}]=e_{1}$}, \mbox{$[e_{3},e_{7}]=e_{1}$}, \mbox{$[e_{4},e_{5}]=- e_{1}$}, \mbox{$[e_{4},e_{7}]=e_{3}$}, \mbox{$[e_{5},e_{6}]=e_{2}$}, \mbox{$[e_{6},e_{7}]=e_{4} + e_{5}$} & $e_{1}$\\\hline
$\mathfrak{n}_{7.126}$\newline {\scriptsize $1,3,4,5,7_{A}$} & \mbox{$[e_{2},e_{6}]=e_{1}$}, \mbox{$[e_{3},e_{7}]=e_{1}$}, \mbox{$[e_{4},e_{7}]=e_{3}$}, \mbox{$[e_{5},e_{7}]=e_{4}$}, \mbox{$[e_{6},e_{7}]=e_{5}$} & $e_{1}$, $- 2 e_{1} e_{4} + e_{3}^{2}$, $3 e_{1}^{2} e_{5} - 3 e_{1} e_{3} e_{4} + e_{3}^{3}$\\\hline
$\mathfrak{n}_{7.127}$\newline {\scriptsize $1,3,4,5,7_{B}$} & \mbox{$[e_{2},e_{6}]=e_{1}$}, \mbox{$[e_{3},e_{7}]=e_{1}$}, \mbox{$[e_{4},e_{7}]=e_{3}$}, \mbox{$[e_{5},e_{6}]=e_{1}$}, \mbox{$[e_{5},e_{7}]=e_{4}$}, \mbox{$[e_{6},e_{7}]=e_{5}$} & $e_{1}$, $- 2 e_{1} e_{4} + e_{3}^{2}$, $- 3 e_{1}^{2} e_{2} + 3 e_{1}^{2} e_{5} - 3 e_{1} e_{3} e_{4} + e_{3}^{3}$\\\hline
$\mathfrak{n}_{7.128}$\newline {\scriptsize $1,3,4,5,7_{C}$} & \mbox{$[e_{2},e_{7}]=e_{1}$}, \mbox{$[e_{3},e_{6}]=e_{1}$}, \mbox{$[e_{4},e_{5}]=- e_{1}$}, \mbox{$[e_{4},e_{7}]=e_{3}$}, \mbox{$[e_{5},e_{7}]=e_{4}$}, \mbox{$[e_{6},e_{7}]=e_{5}$} & $e_{1}$\\\hline
$\mathfrak{n}_{7.129}$\newline {\scriptsize $1,3,4,5,7_{D}$} & \mbox{$[e_{2},e_{6}]=e_{1}$}, \mbox{$[e_{3},e_{7}]=e_{1}$}, \mbox{$[e_{4},e_{6}]=e_{1}$}, \mbox{$[e_{4},e_{7}]=e_{3}$}, \mbox{$[e_{5},e_{6}]=e_{3}$}, \mbox{$[e_{5},e_{7}]=e_{4}$}, \mbox{$[e_{6},e_{7}]=e_{5}$} & $e_{1}$, $2 e_{1} e_{2} - 2 e_{1} e_{4} + e_{3}^{2}$, $3 e_{1}^{2} e_{5} - 3 e_{1} e_{3} e_{4} + e_{3}^{3}$\\\hline
$\mathfrak{n}_{7.130}$\newline {\scriptsize $1,3,4,5,7_{E}$} & \mbox{$[e_{2},e_{7}]=e_{1}$}, \mbox{$[e_{3},e_{6}]=e_{1}$}, \mbox{$[e_{4},e_{5}]=- e_{1}$}, \mbox{$[e_{4},e_{7}]=e_{3}$}, \mbox{$[e_{5},e_{6}]=e_{3}$}, \mbox{$[e_{5},e_{7}]=e_{4}$}, \mbox{$[e_{6},e_{7}]=e_{5}$} & $e_{1}$\\\hline
$\mathfrak{n}_{7.131}$\newline {\scriptsize $1,3,4,5,7_{F}$} & \mbox{$[e_{2},e_{6}]=e_{1}$}, \mbox{$[e_{3},e_{7}]=e_{1}$}, \mbox{$[e_{4},e_{7}]=e_{3}$}, \mbox{$[e_{5},e_{6}]=e_{2}$}, \mbox{$[e_{5},e_{7}]=e_{4}$}, \mbox{$[e_{6},e_{7}]=e_{5}$} & $e_{1}$, $- 2 e_{1} e_{4} + e_{3}^{2}$, $6 e_{1}^{2} e_{5} - 3 e_{1} e_{2}^{2} - 6 e_{1} e_{3} e_{4} + 2 e_{3}^{3}$\\\hline
$\mathfrak{n}_{7.132}$\newline {\scriptsize $1,3,4,5,7_{G}$} & \mbox{$[e_{2},e_{7}]=e_{1}$}, \mbox{$[e_{3},e_{6}]=e_{1}$}, \mbox{$[e_{4},e_{5}]=- e_{1}$}, \mbox{$[e_{4},e_{6}]=e_{1}$}, \mbox{$[e_{4},e_{7}]=e_{3}$}, \mbox{$[e_{5},e_{6}]=e_{2}$}, \mbox{$[e_{5},e_{7}]=e_{4}$}, \mbox{$[e_{6},e_{7}]=e_{5}$} & $e_{1}$\\\hline
$\mathfrak{n}_{7.133}$\newline {\scriptsize $1,3,4,5,7_{I}$} & \mbox{$[e_{2},e_{6}]=e_{1}$}, \mbox{$[e_{2},e_{7}]=e_{1}$}, \mbox{$[e_{3},e_{6}]=e_{1}$}, \mbox{$[e_{4},e_{5}]=- e_{1}$}, \mbox{$[e_{4},e_{6}]=e_{1}$}, \mbox{$[e_{4},e_{7}]=e_{3}$}, \mbox{$[e_{5},e_{6}]=e_{2}$}, \mbox{$[e_{5},e_{7}]=e_{4}$}, \mbox{$[e_{6},e_{7}]=e_{5}$} & $e_{1}$\\\hline
$\mathfrak{n}_{7.134}$\newline {\scriptsize $1,2,4,5,7_{A}$} & \mbox{$[e_{2},e_{7}]=e_{1}$}, \mbox{$[e_{3},e_{5}]=e_{1}$}, \mbox{$[e_{3},e_{6}]=e_{2}$}, \mbox{$[e_{4},e_{7}]=e_{2}$}, \mbox{$[e_{5},e_{7}]=e_{4}$}, \mbox{$[e_{6},e_{7}]=e_{5}$} & $e_{1}$, $- 2 e_{1} e_{4} + e_{2}^{2}$, $2 e_{1} e_{6} - 2 e_{2} e_{5} + e_{4}^{2}$\\\hline
$\mathfrak{n}_{7.135}$\newline {\scriptsize $1,2,4,5,7_{B}$} & \mbox{$[e_{2},e_{7}]=e_{1}$}, \mbox{$[e_{3},e_{5}]=e_{1}$}, \mbox{$[e_{3},e_{6}]=e_{1} + e_{2}$}, \mbox{$[e_{4},e_{7}]=e_{2}$}, \mbox{$[e_{5},e_{7}]=e_{4}$}, \mbox{$[e_{6},e_{7}]=e_{5}$} & $e_{1}$, $- 2 e_{1} e_{4} + e_{2}^{2}$, $6 e_{1}^{2} e_{5} - 6 e_{1}^{2} e_{6} - 6 e_{1} e_{2} e_{4} + 6 e_{1} e_{2} e_{5} - 3 e_{1} e_{4}^{2} + 2 e_{2}^{3}$\\\hline
$\mathfrak{n}_{7.136}$\newline {\scriptsize $1,2,4,5,7_{C}$} & \mbox{$[e_{2},e_{6}]=e_{1}$}, \mbox{$[e_{3},e_{6}]=e_{2}$}, \mbox{$[e_{4},e_{5}]=- e_{1}$}, \mbox{$[e_{4},e_{7}]=e_{2}$}, \mbox{$[e_{5},e_{7}]=e_{4}$}, \mbox{$[e_{6},e_{7}]=e_{5}$} & $e_{1}$, $- 2 e_{1} e_{3} + e_{2}^{2}$, $- 2 e_{1} e_{7} - 2 e_{2} e_{5} + e_{4}^{2}$\\\hline
$\mathfrak{n}_{7.137}$\newline {\scriptsize $1,2,4,5,7_{D}$} & \mbox{$[e_{2},e_{6}]=e_{1}$}, \mbox{$[e_{3},e_{6}]=e_{2}$}, \mbox{$[e_{3},e_{7}]=e_{1}$}, \mbox{$[e_{4},e_{5}]=- e_{1}$}, \mbox{$[e_{4},e_{7}]=e_{2}$}, \mbox{$[e_{5},e_{7}]=e_{4}$}, \mbox{$[e_{6},e_{7}]=e_{5}$} & $e_{1}$\\\hline
$\mathfrak{n}_{7.138}$\newline {\scriptsize $1,2,4,5,7_{E}$} & \mbox{$[e_{2},e_{7}]=e_{1}$}, \mbox{$[e_{3},e_{5}]=e_{1}$}, \mbox{$[e_{3},e_{6}]=e_{2}$}, \mbox{$[e_{4},e_{6}]=e_{1}$}, \mbox{$[e_{4},e_{7}]=e_{2}$}, \mbox{$[e_{5},e_{6}]=e_{2}$}, \mbox{$[e_{5},e_{7}]=e_{4}$}, \mbox{$[e_{6},e_{7}]=e_{5}$} & $e_{1}$\\\hline
$\mathfrak{n}_{7.139}$\newline {\scriptsize $1,2,4,5,7_{F}$} & \mbox{$[e_{2},e_{6}]=e_{1}$}, \mbox{$[e_{3},e_{6}]=e_{2}$}, \mbox{$[e_{4},e_{5}]=- e_{1}$}, \mbox{$[e_{4},e_{7}]=e_{2}$}, \mbox{$[e_{5},e_{6}]=e_{2}$}, \mbox{$[e_{5},e_{7}]=e_{4}$}, \mbox{$[e_{6},e_{7}]=e_{5}$} & $e_{1}$, $- 2 e_{1} e_{3} + e_{2}^{2}$, $- 6 e_{1}^{2} e_{7} - 6 e_{1} e_{2} e_{5} + 3 e_{1} e_{4}^{2} + 2 e_{2}^{3}$\\\hline
$\mathfrak{n}_{7.140}$\newline {\scriptsize $1,2,4,5,7_{G}$} & \mbox{$[e_{2},e_{6}]=e_{1}$}, \mbox{$[e_{3},e_{6}]=e_{2}$}, \mbox{$[e_{3},e_{7}]=e_{1}$}, \mbox{$[e_{4},e_{5}]=- e_{1}$}, \mbox{$[e_{4},e_{7}]=e_{2}$}, \mbox{$[e_{5},e_{6}]=e_{2}$}, \mbox{$[e_{5},e_{7}]=e_{4}$}, \mbox{$[e_{6},e_{7}]=e_{5}$} & $e_{1}$\\\hline
$\mathfrak{n}_{7.141}$\newline {\scriptsize $1,2,4,5,7_{H}$} & \mbox{$[e_{2},e_{7}]=e_{1}$}, \mbox{$[e_{3},e_{7}]=e_{2}$}, \mbox{$[e_{4},e_{5}]=e_{1}$}, \mbox{$[e_{4},e_{6}]=e_{2}$}, \mbox{$[e_{5},e_{6}]=e_{3}$}, \mbox{$[e_{5},e_{7}]=e_{4}$}, \mbox{$[e_{6},e_{7}]=e_{5}$} & $e_{1}$, $- 2 e_{1} e_{3} + e_{2}^{2}$, $e_{1} e_{6} - e_{2} e_{5} + e_{3} e_{4}$\\\hline
$\mathfrak{n}_{7.142}$\newline {\scriptsize $1,2,4,5,7_{I}$} & \mbox{$[e_{2},e_{7}]=e_{1}$}, \mbox{$[e_{3},e_{6}]=e_{1}$}, \mbox{$[e_{3},e_{7}]=e_{2}$}, \mbox{$[e_{4},e_{5}]=e_{1}$}, \mbox{$[e_{4},e_{6}]=e_{2}$}, \mbox{$[e_{5},e_{6}]=e_{3}$}, \mbox{$[e_{5},e_{7}]=e_{4}$}, \mbox{$[e_{6},e_{7}]=e_{5}$} & $e_{1}$\\\hline
$\mathfrak{n}_{7.143}$\newline {\scriptsize $1,2,4,5,7_{J}$} & \mbox{$[e_{2},e_{7}]=e_{1}$}, \mbox{$[e_{3},e_{6}]=e_{1}$}, \mbox{$[e_{3},e_{7}]=e_{2}$}, \mbox{$[e_{4},e_{5}]=e_{1}$}, \mbox{$[e_{4},e_{6}]=e_{2}$}, \mbox{$[e_{4},e_{7}]=e_{1}$}, \mbox{$[e_{5},e_{6}]=e_{3}$}, \mbox{$[e_{5},e_{7}]=e_{4}$}, \mbox{$[e_{6},e_{7}]=e_{5}$} & $e_{1}$\\\hline
$\mathfrak{n}_{7.144}$\newline {\scriptsize $1,2,4,5,7_{K}$} & \mbox{$[e_{2},e_{7}]=e_{1}$}, \mbox{$[e_{3},e_{7}]=e_{2}$}, \mbox{$[e_{4},e_{5}]=e_{1}$}, \mbox{$[e_{4},e_{6}]=e_{2}$}, \mbox{$[e_{4},e_{7}]=e_{1}$}, \mbox{$[e_{5},e_{6}]=e_{3}$}, \mbox{$[e_{5},e_{7}]=e_{4}$}, \mbox{$[e_{6},e_{7}]=e_{5}$} & $e_{1}$, $- 2 e_{1} e_{3} + e_{2}^{2}$, $3 e_{1}^{2} e_{6} - 3 e_{1} e_{2} e_{3} - 3 e_{1} e_{2} e_{5} + 3 e_{1} e_{3} e_{4} + e_{2}^{3}$\\\hline
$\mathfrak{n}_{7.145}$\newline {\scriptsize $1,2,4,5,7_{L}$} & \mbox{$[e_{2},e_{6}]=e_{1}$}, \mbox{$[e_{2},e_{7}]=e_{1}$}, \mbox{$[e_{3},e_{5}]=- e_{1}$}, \mbox{$[e_{3},e_{7}]=e_{2}$}, \mbox{$[e_{4},e_{5}]=e_{1}$}, \mbox{$[e_{4},e_{6}]=e_{2}$}, \mbox{$[e_{5},e_{6}]=e_{3}$}, \mbox{$[e_{5},e_{7}]=e_{4}$}, \mbox{$[e_{6},e_{7}]=e_{5}$} & $e_{1}$, $- 2 e_{1} e_{3} - 2 e_{1} e_{4} + e_{2}^{2}$, $e_{1} e_{6} - e_{1} e_{7} - e_{2} e_{5} + e_{3} e_{4}$\\\hline
$\mathfrak{n}_{7.146}$\newline {\scriptsize $1,2,4,5,7_{M}$} & \mbox{$[e_{2},e_{6}]=e_{1}$}, \mbox{$[e_{2},e_{7}]=e_{1}$}, \mbox{$[e_{3},e_{5}]=- e_{1}$}, \mbox{$[e_{3},e_{7}]=e_{2}$}, \mbox{$[e_{4},e_{5}]=e_{1}$}, \mbox{$[e_{4},e_{6}]=e_{2}$}, \mbox{$[e_{4},e_{7}]=e_{1}$}, \mbox{$[e_{5},e_{6}]=e_{3}$}, \mbox{$[e_{5},e_{7}]=e_{4}$}, \mbox{$[e_{6},e_{7}]=e_{5}$} & $e_{1}$\\\hline

$\mathfrak{n}_{7.147}(\xi)$\newline {\scriptsize $1,2,4,5,7_{N}$}\newline {\scriptsize $\xi\ne0$} & \mbox{$[e_{2},e_{6}]=e_{1}$}, \mbox{$[e_{2},e_{7}]=e_{1}$}, \mbox{$[e_{3},e_{5}]=- e_{1}$}, \mbox{$[e_{3},e_{6}]=\xi e_{1}$}, \mbox{$[e_{3},e_{7}]=e_{2}$}, \mbox{$[e_{4},e_{5}]=e_{1}$}, \mbox{$[e_{4},e_{6}]=e_{2}$}, \mbox{$[e_{4},e_{7}]=e_{1}$}, \mbox{$[e_{5},e_{6}]=e_{3}$}, \mbox{$[e_{5},e_{7}]=e_{4}$}, \mbox{$[e_{6},e_{7}]=e_{5}$} & $e_{1}$ ($\xi\ne 1$);\newline $e_{1}$, $2 e_{1} e_{2} - 2 e_{1} e_{3} - 2 e_{1} e_{4} + e_{2}^{2}$, $- 3 e_{1}^{2} e_{2} + 3 e_{1}^{2} e_{3} + 3 e_{1}^{2} e_{4} + 3 e_{1}^{2} e_{5} + 3 e_{1}^{2} e_{6} - 3 e_{1}^{2} e_{7} - 3 e_{1} e_{2} e_{3} - 3 e_{1} e_{2} e_{4} - 3 e_{1} e_{2} e_{5} + 3 e_{1} e_{3} e_{4} + e_{2}^{3}$ ($\xi=1$)\\\hline
$\mathfrak{n}_{7.148}$\newline {\scriptsize $1,2,3,5,7_{A}$} & \mbox{$[e_{2},e_{7}]=e_{1}$}, \mbox{$[e_{3},e_{4}]=- e_{1}$}, \mbox{$[e_{3},e_{7}]=e_{2}$}, \mbox{$[e_{4},e_{5}]=e_{2}$}, \mbox{$[e_{4},e_{6}]=e_{3}$}, \mbox{$[e_{5},e_{7}]=e_{3}$}, \mbox{$[e_{6},e_{7}]=e_{5}$} & $e_{1}$, $3 e_{1}^{2} e_{5} - 3 e_{1} e_{2} e_{3} + e_{2}^{3}$, $- 4 e_{1}^{3} e_{6} + 4 e_{1}^{2} e_{2} e_{5} + 2 e_{1}^{2} e_{3}^{2} - 4 e_{1} e_{2}^{2} e_{3} + e_{2}^{4}$\\\hline
$\mathfrak{n}_{7.149}$\newline {\scriptsize $1,2,3,5,7_{B}$} & \mbox{$[e_{2},e_{7}]=e_{1}$}, \mbox{$[e_{3},e_{4}]=- e_{1}$}, \mbox{$[e_{3},e_{7}]=e_{2}$}, \mbox{$[e_{4},e_{5}]=e_{2}$}, \mbox{$[e_{4},e_{6}]=e_{1} + e_{3}$}, \mbox{$[e_{5},e_{7}]=e_{3}$}, \mbox{$[e_{6},e_{7}]=e_{5}$} & $e_{1}$, $3 e_{1}^{2} e_{5} - 3 e_{1} e_{2} e_{3} + e_{2}^{3}$, $4 e_{1}^{3} e_{3} - 4 e_{1}^{3} e_{6} - 2 e_{1}^{2} e_{2}^{2} + 4 e_{1}^{2} e_{2} e_{5} + 2 e_{1}^{2} e_{3}^{2} - 4 e_{1} e_{2}^{2} e_{3} + e_{2}^{4}$\\\hline
$\mathfrak{n}_{7.150}$\newline {\scriptsize $1,2,3,5,7_{C}$} & \mbox{$[e_{2},e_{7}]=e_{1}$}, \mbox{$[e_{3},e_{4}]=- e_{1}$}, \mbox{$[e_{3},e_{7}]=e_{2}$}, \mbox{$[e_{4},e_{5}]=e_{2}$}, \mbox{$[e_{4},e_{6}]=e_{3}$}, \mbox{$[e_{5},e_{6}]=e_{1}$}, \mbox{$[e_{5},e_{7}]=e_{3}$}, \mbox{$[e_{6},e_{7}]=e_{5}$} & $e_{1}$\\\hline
$\mathfrak{n}_{7.151}$\newline {\scriptsize $1,2,3,4,5,7_{A}$} & \mbox{$[e_{2},e_{7}]=e_{1}$}, \mbox{$[e_{3},e_{7}]=e_{2}$}, \mbox{$[e_{4},e_{7}]=e_{3}$}, \mbox{$[e_{5},e_{7}]=e_{4}$}, \mbox{$[e_{6},e_{7}]=e_{5}$} & $e_{1}$, $- 2 e_{1} e_{3} + e_{2}^{2}$, $2 e_{1} e_{5} - 2 e_{2} e_{4} + e_{3}^{2}$, $3 e_{1}^{2} e_{4} - 3 e_{1} e_{2} e_{3} + e_{2}^{3}$, $- 5 e_{1}^{2} e_{6} + 5 e_{1} e_{2} e_{5} - e_{1} e_{3} e_{4} - 2 e_{2}^{2} e_{4} + e_{2} e_{3}^{2}$\\\hline
$\mathfrak{n}_{7.152}$\newline {\scriptsize $1,2,3,4,5,7_{B}$} & \mbox{$[e_{2},e_{7}]=e_{1}$}, \mbox{$[e_{3},e_{7}]=e_{2}$}, \mbox{$[e_{4},e_{7}]=e_{3}$}, \mbox{$[e_{5},e_{6}]=e_{1}$}, \mbox{$[e_{5},e_{7}]=e_{4}$}, \mbox{$[e_{6},e_{7}]=e_{5}$} & $e_{1}$, $- 2 e_{1} e_{3} + e_{2}^{2}$, $3 e_{1}^{2} e_{4} - 3 e_{1} e_{2} e_{3} + e_{2}^{3}$\\\hline
$\mathfrak{n}_{7.153}$\newline {\scriptsize $1,2,3,4,5,7_{C}$} & \mbox{$[e_{2},e_{7}]=e_{1}$}, \mbox{$[e_{3},e_{6}]=e_{1}$}, \mbox{$[e_{3},e_{7}]=e_{2}$}, \mbox{$[e_{4},e_{5}]=- e_{1}$}, \mbox{$[e_{4},e_{7}]=e_{3}$}, \mbox{$[e_{5},e_{7}]=e_{4}$}, \mbox{$[e_{6},e_{7}]=e_{5}$} & $e_{1}$\\\hline
$\mathfrak{n}_{7.154}$\newline {\scriptsize $1,2,3,4,5,7_{D}$} & \mbox{$[e_{2},e_{7}]=e_{1}$}, \mbox{$[e_{3},e_{7}]=e_{2}$}, \mbox{$[e_{4},e_{6}]=e_{1}$}, \mbox{$[e_{4},e_{7}]=e_{3}$}, \mbox{$[e_{5},e_{6}]=e_{2}$}, \mbox{$[e_{5},e_{7}]=e_{4}$}, \mbox{$[e_{6},e_{7}]=e_{5}$} & $e_{1}$, $- 2 e_{1} e_{3} + e_{2}^{2}$, $2 e_{1} e_{5} - 2 e_{2} e_{4} + e_{3}^{2}$\\\hline
$\mathfrak{n}_{7.155}$\newline {\scriptsize $1,2,3,4,5,7_{E}$} & \mbox{$[e_{2},e_{7}]=e_{1}$}, \mbox{$[e_{3},e_{7}]=e_{2}$}, \mbox{$[e_{4},e_{6}]=e_{1}$}, \mbox{$[e_{4},e_{7}]=e_{3}$}, \mbox{$[e_{5},e_{6}]=e_{1} + e_{2}$}, \mbox{$[e_{5},e_{7}]=e_{4}$}, \mbox{$[e_{6},e_{7}]=e_{5}$} & $e_{1}$, $- 2 e_{1} e_{3} + e_{2}^{2}$, $6 e_{1}^{2} e_{4} - 6 e_{1}^{2} e_{5} - 6 e_{1} e_{2} e_{3} + 6 e_{1} e_{2} e_{4} - 3 e_{1} e_{3}^{2} + 2 e_{2}^{3}$\\\hline
$\mathfrak{n}_{7.156}$\newline {\scriptsize $1,2,3,4,5,7_{F}$} & \mbox{$[e_{2},e_{7}]=e_{1}$}, \mbox{$[e_{3},e_{6}]=e_{1}$}, \mbox{$[e_{3},e_{7}]=e_{2}$}, \mbox{$[e_{4},e_{5}]=- e_{1}$}, \mbox{$[e_{4},e_{6}]=e_{1}$}, \mbox{$[e_{4},e_{7}]=e_{3}$}, \mbox{$[e_{5},e_{6}]=e_{2}$}, \mbox{$[e_{5},e_{7}]=e_{4}$}, \mbox{$[e_{6},e_{7}]=e_{5}$} & $e_{1}$\\\hline
$\mathfrak{n}_{7.157}$\newline {\scriptsize $1,2,3,4,5,7_{G}$} & \mbox{$[e_{2},e_{7}]=e_{1}$}, \mbox{$[e_{3},e_{6}]=e_{1}$}, \mbox{$[e_{3},e_{7}]=e_{2}$}, \mbox{$[e_{4},e_{6}]=e_{2}$}, \mbox{$[e_{4},e_{7}]=e_{3}$}, \mbox{$[e_{5},e_{6}]=e_{3}$}, \mbox{$[e_{5},e_{7}]=e_{4}$}, \mbox{$[e_{6},e_{7}]=e_{5}$} & $e_{1}$, $3 e_{1}^{2} e_{4} - 3 e_{1} e_{2} e_{3} + e_{2}^{3}$, $- 4 e_{1}^{3} e_{5} + 4 e_{1}^{2} e_{2} e_{4} + 2 e_{1}^{2} e_{3}^{2} - 4 e_{1} e_{2}^{2} e_{3} + e_{2}^{4}$\\\hline
$\mathfrak{n}_{7.158}$\newline {\scriptsize $1,2,3,4,5,7_{H}$} & \mbox{$[e_{2},e_{7}]=e_{1}$}, \mbox{$[e_{3},e_{6}]=e_{1}$}, \mbox{$[e_{3},e_{7}]=e_{2}$}, \mbox{$[e_{4},e_{6}]=e_{2}$}, \mbox{$[e_{4},e_{7}]=e_{3}$}, \mbox{$[e_{5},e_{6}]=e_{1} + e_{3}$}, \mbox{$[e_{5},e_{7}]=e_{4}$}, \mbox{$[e_{6},e_{7}]=e_{5}$} & $e_{1}$, $3 e_{1}^{2} e_{4} - 3 e_{1} e_{2} e_{3} + e_{2}^{3}$, $4 e_{1}^{3} e_{3} - 4 e_{1}^{3} e_{5} - 2 e_{1}^{2} e_{2}^{2} + 4 e_{1}^{2} e_{2} e_{4} + 2 e_{1}^{2} e_{3}^{2} - 4 e_{1} e_{2}^{2} e_{3} + e_{2}^{4}$\\\hline

$\mathfrak{n}_{7.159}(\xi)$\newline {\scriptsize $1,2,3,4,5,7_{I}$}\newline {\scriptsize $\xi\ne1$} & \mbox{$[e_{2},e_{7}]=e_{1}$}, \mbox{$[e_{3},e_{6}]=\xi e_{1}$}, \mbox{$[e_{3},e_{7}]=e_{2}$}, \mbox{$[e_{4},e_{5}]=(1-\xi)e_{1}$}, \mbox{$[e_{4},e_{6}]=e_{2}$}, \mbox{$[e_{4},e_{7}]=e_{3}$}, \mbox{$[e_{5},e_{6}]=e_{3}$}, \mbox{$[e_{5},e_{7}]=e_{4}$}, \mbox{$[e_{6},e_{7}]=e_{5}$} & $e_{1}$ ($\xi\ne0$);\newline $e_{1}$, $- 2 e_{1} e_{3} + e_{2}^{2}$, $- 15 e_{1}^{4} e_{6} + 15 e_{1}^{3} e_{2} e_{5} - 15 e_{1}^{3} e_{3} e_{4} + 15 e_{1}^{2} e_{2} e_{3}^{2} - 10 e_{1} e_{2}^{3} e_{3} + 2 e_{2}^{5}$ ($\xi=0$)\\\hline
\end{longtable}\endgroup

\medskip\noindent\textbf{Remark on the relation to Seeley's classification.}
We use the ordered basis~$(a,b,c,\allowbreak d,e,f,g)$ of~\cite[pp.~482--494]{seeley1993},
with the corrections described above. For a decomposable entry, letters
absent from the brackets in~\cite{seeley1993} complete the basis by central
vectors. The tuple~$(m_1,\ldots,m_7)$ below gives the images of Seeley's
basis vectors in the basis of~$\mathfrak n_{7.k}$ from Table~\ref{tab:casimir}.
Its $j$th component denotes~$\operatorname{sgn}(m_j)e_{|m_j|}$.
In particular, rows~94--159, the 66 entries with a one-dimensional
center, all use the reversed basis with every sign changed.

\begin{center}
\small
\begin{tabular}{@{}ll@{}}
\textbf{Row numbers $k$} & \textbf{Signed indices of the images} \\
\hline
0, 1, 3--6, 8, 12, 94--159 & $(-7,-6,-5,-4,-3,-2,-1)$ \\
2 & $(5,6,7,3,4,1,2)$ \\
7, 83, 90 & $(6,7,5,-4,3,-2,-1)$ \\
9, 10, 26--30, 32--40 & $(6,7,5,4,3,-2,-1)$ \\
11 & $(6,7,4,5,3,-1,-2)$ \\
13 & $(6,7,4,5,-3,1,2)$ \\
14--18, 41, 42, 48--60 & $(6,7,5,4,-3,-2,-1)$ \\
19--21 & $(6,7,5,4,-3,-1,-2)$ \\
22--24, 61, 62, 65, 68, 71--73, 75, 84, 86, 88, 92 & $(6,-7,-5,4,-3,-2,-1)$ \\
25 & $(6,7,3,4,-5,1,2)$ \\
31 & $(6,7,5,3,-4,-2,1)$ \\
43--46 & $(6,-7,5,-4,-3,2,1)$ \\
47 & $(6,5,-7,-3,-4,-2,1)$ \\
63, 76, 82, 85, 87, 89, 91, 93 & $(6,-7,-5,4,-3,2,-1)$ \\
64, 69, 70, 77, 79, 80 & $(6,-7,-5,4,3,-2,-1)$ \\
66, 67 & $(6,-7,-5,4,-3,-2,1)$ \\
74 & $(-6,7,-5,-4,3,2,-1)$ \\
78, 81 & $(6,7,5,4,-3,2,-1)$ \\
\end{tabular}
\end{center}

For instance, in row~81 the image of Seeley's bracket~$[b,d]=e+g$ is
$[e_7,e_4]=-e_3-e_1$. The bracket~$[c,d]=f$ becomes~$[e_5,e_4]=e_2$.

At four excluded parameter values, the brackets already coincide in the
displayed bases:
\[
\begin{aligned}
\mathfrak n_{7.100}(\tfrac12)&=\mathfrak n_{7.98},&
\mathfrak n_{7.119}(\tfrac12)&=\mathfrak n_{7.117},\\
\mathfrak n_{7.147}(0)&=\mathfrak n_{7.146},&
\mathfrak n_{7.159}(1)&=\mathfrak n_{7.157}.
\end{aligned}
\]
For the other excluded values, the following tuples give the images of
the ordered basis~$(e_1,\ldots,e_7)$ of the source in the basis of the target.
The last row gives the isomorphism from the isolated entry to the specialization.

\begin{center}
\small
\begin{tabular}{@{}lll@{}}
\textbf{Source} & \textbf{Target} & \textbf{Images of the source basis} \\
\hline
$\mathfrak n_{7.100}(0)$ & $\mathfrak n_{7.58}$ & $(e_1,e_2,e_3,-e_4,-e_5,-e_6,e_7)$ \\
$\mathfrak n_{7.100}(1)$ & $\mathfrak n_{7.58}$ & $(e_1,e_3,e_4,-e_2,-e_6,-e_7,-e_5)$ \\
$\mathfrak n_{7.100}(-1)$ & $\mathfrak n_{7.98}$ & $(-\tfrac12 e_1,e_4,e_2,e_3,e_7,e_5,e_6)$ \\
$\mathfrak n_{7.100}(2)$ & $\mathfrak n_{7.98}$ & $(\tfrac12 e_1,-e_3,-e_2,-e_4,e_5,e_7,e_6)$ \\
$\mathfrak n_{7.119}(0)$ & $\mathfrak n_{7.79}$ & $(e_2,e_1,-e_3,-e_4,e_5,e_7,-e_6)$ \\
$\mathfrak n_{7.81}$ & $\mathfrak n_{7.125}(1)$ & $(e_3-e_2,e_1,e_2,e_4,e_4+e_5,e_6,e_7-e_6)$
\end{tabular}
\end{center}

\section{Conclusion}

We have presented polynomial functional bases of invariants for Seeley's classification as corrected by Gong~\cite{gong1998,seeley1993}, including all admissible values in the six one-parameter families. The chosen~$K$-canonical bases satisfy~$[\mathfrak g,A_j]\subseteq A_{j-1}$, making the matrices of inner automorphisms upper unitriangular and facilitating successive elimination of the group parameters. The computations have been illustrated by several examples, and Table~\ref{tab:casimir} lists the resulting Casimir operators. The symbolic moving-frame calculations and independent checks were performed with SymPy~\cite{sympy}; Maple~\cite{maple} was used in the earlier computations. Successive normalization gives a polynomial functional basis in every case, including the exceptional parameter values. The compact expressions in the table are recovered by invariantizing their restrictions to the cross-section. The infinitesimal equations~\eqref{SystemForLieAlgebraInvariants} and exact rank and Jacobian computations were used only for independent verification.
The table complements the descriptions of the full polynomial invariant algebras in~\cite{Ooms2009,Ooms2012} by presenting explicit functional bases in the chosen coordinates and showing how to construct them by successive normalization.

\medskip

\textbf{Acknowledgments.}
The authors are grateful to Roman Popovych and Oleksandra Vinnichenko
for helpful discussions and suggestions.

\medskip

\textbf{Funding.} V.M.B. gratefully acknowledges the support of the National Research Foundation of Ukraine under project 2025.07/0405, \emph{Algebraic methods for the study of equations of mathematical physics}. The content of this paper is the sole responsibility of the Institute of Mathematics of the National Academy of Sciences of Ukraine. The views expressed herein do not necessarily reflect those of the National Research Foundation of Ukraine.


\begingroup\footnotesize\begin{thebibliography}{99}\setlength{\itemsep}{0pt}\setlength{\parsep}{0pt}
\addcontentsline{toc}{section}{References}
\providecommand{\eprint}[2][]{\href{https://arxiv.org/abs/#2}{arXiv:#2}}

\bibitem{Abellanas-Martinez}
Abellanas L., Mart\'inez Alonso L., \href{https://doi.org/10.1063/1.522727}{A general setting for Casimir invariants}, \emph{J. Math. Phys.} \textbf{16} (1975), 1580--1584.

\bibitem{Abellanas-Martinez2}
Abellanas L., Mart\'inez Alonso L., \href{https://doi.org/10.1063/1.524080}{Invariants in enveloping algebras under the action of Lie algebras of derivations}, \emph{J. Math. Phys.} \textbf{20} (1979), 437--440.

\bibitem{beltrametti}
Beltrametti E.G., Blasi A., \href{https://doi.org/10.1016/0031-9163(66)91048-1}{On the number of Casimir operators associated with any Lie group}, \emph{Phys. Lett.} \textbf{20} (1966), 62--64.

\bibitem{Boyko2006}
Boyko V.M., Patera J., Popovych R.O., \href{https://doi.org/10.1088/0305-4470/39/20/009}{Computation of invariants of Lie algebras by means of moving frames}, \emph{J. Phys. A: Math. Gen.} \textbf{39} (2006), 5749--5762, extended and revised version, \eprint{math-ph/0602046}.

\bibitem{Boyko2007a}
Boyko V.M., Patera J., Popovych R.O., \href{https://doi.org/10.1088/1751-8113/40/1/006}{Invariants of Lie algebras with fixed structure of nilradicals}, \emph{J. Phys. A: Math. Theor.} \textbf{40} (2007), 113--130, \eprint{math-ph/0606045}.

\bibitem{Boyko2007b}
Boyko V.M., Patera J., Popovych R.O., \href{https://doi.org/10.1088/1751-8113/40/27/009}{Invariants of triangular Lie algebras}, \emph{J. Phys. A: Math. Theor.} \textbf{40} (2007), 7557--7572, \eprint{0704.0937}.

\bibitem{Boyko2007c}
Boyko V.M., Patera J., Popovych R.O., \href{https://doi.org/10.1088/1751-8113/40/32/005}{Invariants of triangular Lie algebras with one nil-independent diagonal element}, \emph{J. Phys. A: Math. Theor.} \textbf{40} (2007), 9783--9792, \eprint{0705.2394}.

\bibitem{Boyko2008}
Boyko V.M., Patera J., Popovych R.O., \href{https://doi.org/10.1016/j.laa.2007.08.017}{Invariants of solvable Lie algebras with triangular nilradicals and diagonal nilindependent elements}, \emph{Linear Algebra Appl.} \textbf{428} (2008), 834--854, \eprint{0706.2465}.

\bibitem{Boyko&Patera&Popovych2009}
Boyko V.M., Patera J., Popovych R.O., Invariants of Lie algebras via moving frames, in Proceedings of the Fourth International Workshop ``Group Analysis of Differential Equations and Integrable Systems'' (October 26--30, 2008, Protaras, Cyprus), University of Cyprus, Nicosia, 2009, 36--44, \eprint{0904.4462}.

\bibitem{Campoamor-Rausch}
Campoamor-Stursberg R., Rausch de Traubenberg M., \href{https://doi.org/10.1142/11081}{Group theory in physics: A practitioner's guide}, World Scientific, Singapore, 2018.

\bibitem{Casimir}
Casimir H.B.G., \"Uber die Konstruktion einer zu den irreduzibelen Dar\-stel\-lungen halbeinfacher kontinuierlicher Gruppen geh\"origen Differentialgleichung, \emph{Proc. R. Acad. Amsterdam} \textbf{34} (1931), 844--846.

\bibitem{Dixmier}
Dixmier J., Enveloping algebras, North-Holland Math. Library, Vol.~14, North-Holland, Amsterdam, 1977.

\bibitem{Dixmier1957}
Dixmier J., \href{https://doi.org/10.24033/bsmf.1492}{Sur les repr\'esentations unitaires des groupes de Lie nilpotents. II}, \emph{Bull. Soc. Math. France} \textbf{85} (1957), 325--388.

\bibitem{Fels-Olver}
Fels M.E., Olver P.J., \href{https://doi.org/10.1023/A:1005878210297}{Moving coframes. I. A practical algorithm}, \emph{Acta Appl. Math.} \textbf{51} (1998), 161--213.

\bibitem{Fels-Olver2}
Fels M.E., Olver P.J., \href{https://doi.org/10.1023/A:1006195823000}{Moving coframes. II. Regularization and theoretical foundations}, \emph{Acta Appl. Math.} \textbf{55} (1999), 127--208.

\bibitem{gong1998}
Gong M.-P., Classification of nilpotent Lie algebras of dimension 7 (over algebraically closed fields and~$\mathbb R$), Ph.D. Thesis, University of Waterloo, 1998, \url{https://uwspace.uwaterloo.ca/items/5d5e1909-9990-4903-90ad-d14e4e74cd6f}.

\bibitem{Linton-Holderbaum}
Linton C., Holderbaum W., Biggs J., \href{https://www.ejpam.com/ejpam/article/view/1515}{Casimirs and {L}ax operators from the structure of {L}ie algebras}, \emph{Eur. J. Pure Appl. Math.} \textbf{5} (2012), 567--583.

\bibitem{Magnin2010}
Magnin L., \href{https://doi.org/10.1007/s10468-009-9172-3}{Determination of 7-dimensional indecomposable nilpotent complex Lie algebras by adjoining a derivation to 6-dimensional Lie algebras}, \emph{Algebr. Represent. Theory} \textbf{13} (2010), 723--753.

\bibitem{maple}
Maplesoft, Maple 2023 (computer software), Waterloo, Ontario, 2023, \url{https://www.maplesoft.com/documentation_center/}.

\bibitem{sympy}
Meurer A., Smith C.P., Paprocki M., et al., SymPy: symbolic computing in Python, \emph{PeerJ Comput. Sci.} \textbf{3} (2017), e103, \href{https://doi.org/10.7717/peerj-cs.103}{doi:10.7717/peerj-cs.103}.

\bibitem{morozov1958}
Morozov V.V., \href{https://www.mathnet.ru/eng/ivm/y1958/i4/p161}{Classification of nilpotent Lie algebras of sixth order}, \emph{Izv. Vyssh. Uchebn. Zaved. Mat.} (1958), no.~4 (5), 161--171.

\bibitem{mubarakzyanov1963.1}
Mubarakzyanov G.M., \href{https://www.mathnet.ru/eng/ivm/y1963/i1/p114}{On solvable Lie algebras}, \emph{Izv. Vyssh. Uchebn. Zaved. Mat.}  (1963), no.~1 (32), 114--123.

\bibitem{Ndogmo}
Ndogmo J.C., \href{https://doi.org/10.4153/CMB-2011-102-2}{Casimir operators and nilpotent radicals}, \emph{Canad. Math. Bull.} \textbf{55} (2012), 579--585.

\bibitem{Ooms2009}
Ooms A.I., \href{https://doi.org/10.1016/j.jalgebra.2008.10.026}{Computing invariants and semi-invariants by means of Frobenius Lie algebras}, \emph{J. Algebra} \textbf{321} (2009), 1293--1312, \eprint{0806.4178}.

\bibitem{Ooms2012}
Ooms A.I., \href{https://doi.org/10.1016/j.jalgebra.2012.04.029}{The Poisson center and polynomial, maximal Poisson commutative subalgebras, especially for nilpotent Lie algebras of dimension at most seven}, \emph{J. Algebra} \textbf{365} (2012), 83--113, \eprint{1110.0363}.

\bibitem{patera-sharp}
Patera J., Sharp R.T., Winternitz P., Zassenhaus H., \href{https://doi.org/10.1063/1.522992}{Invariants of real low dimension Lie algebras}, \emph{J. Math. Phys.} \textbf{17} (1976), 986--994.

\bibitem{Pecina-Cruz}
Pecina-Cruz J.N., \href{https://doi.org/10.1063/1.530458}{An algorithm to calculate the invariants of any Lie algebra}, \emph{J. Math. Phys.} \textbf{35} (1994), 3146--3162.

\bibitem{popovych2003}
Popovych R.O., Boyko V.M., Nesterenko M.O., Lutfullin M.W., \href{https://doi.org/10.1088/0305-4470/36/26/309}{Realizations of real low-dimensional Lie algebras}, \emph{J. Phys. A: Math. Gen.} \textbf{36} (2003), 7337--7360, extended and revised version, \eprint{math-ph/0301029}.

\bibitem{Romdhani}
Romdhani M., \href{https://doi.org/10.1080/03081088908817910}{Classification of real and complex nilpotent Lie algebras of dimension 7}, \emph{Linear Multilinear Algebra} \textbf{24} (1989), 167--189.

\bibitem{Salgado-Gonzalez}
Salgado-Gonz\'alez G., \href{https://doi.org/10.1016/j.geomphys.2019.06.014}{Invariants of contact Lie algebras}, \emph{J. Geom. Phys.} \textbf{144} (2019), 388--396.

\bibitem{seeley1993}
Seeley C., \href{https://doi.org/10.2307/2154390}{7-dimensional nilpotent Lie algebras}, \emph{Trans. Amer. Math. Soc.} \textbf{335} (1993), 479--496.

\bibitem{turkowski1990}
Turkowski P., \href{https://doi.org/10.1063/1.528721}{Solvable Lie algebras of dimension six}, \emph{J. Math. Phys.} \textbf{31} (1990), 1344--1350.

\bibitem{Vorushilov}
Vorushilov K.S., \href{https://www.mathnet.ru/eng/sm9501}{Complete sets of polynomials in bi-involution on nilpotent seven-dimensional Lie algebras}, \emph{Sb. Math.} \textbf{212} (2021), 1193--1207.

\end{thebibliography}\endgroup
\end{document}